\documentclass{article}

\usepackage{graphicx, pdfpages}
\usepackage{natbib}
\usepackage{latexsym}
\usepackage{mathtools}
\usepackage{amsmath, amsthm, amssymb}

\newtheorem{Prop}{Proposition}
     
\newcommand{\be}{\begin{equation}}
\newcommand{\ee}{\end{equation}} 
\newcommand{\lb}{\label}

\newcommand{\OL}{\overline}

\newcommand{\const}{({\rm const.})}

\newcommand{\ba}{{\bf a}}

\newcommand{\bk}{{\bf k}}

\newcommand{\bp}{{\bf p}}
\newcommand{\bq}{{\bf q}}
\newcommand{\br}{{\bf r}}
\newcommand{\bu}{{\bf u}}

\newcommand{\bx}{{\bf x}}

\newcommand{\bJ}{{\bf J}}

\newcommand{\bs}{{\bf s}}

\newcommand{\wt}{\widetilde}

\newcommand{\grad}{{\mbox{\boldmath $\nabla$}}}
\newcommand{\bdot}{{\mbox{\boldmath $\cdot$}}}

\newcommand{\bzed}{{\mbox{\boldmath $0$}}}
\newcommand{\boell}{{\mbox{\boldmath $\ell$}}}

\begin{document}

\baselineskip=18pt
\begin{center}
\begin{LARGE}
{\bf Scale decomposition in compressible turbulence}\\
\end{LARGE}

\bigskip
\bigskip

Hussein Aluie\\
{\it
Applied Mathematics and Plasma Physics (T-5),\\
Computer, Computational, and Statistical Sciences (CCS-2),\\
\& Center for Non-linear Studies,\\ Los Alamos National Laboratory, MS-B258
Los Alamos, NM 87545, USA}

\bigskip
\bigskip

\begin{abstract}
This work presents a rigorous framework based on coarse-graining to analyze 
highly compressible turbulence. 
We show how the requirement that viscous effects on the dynamics of 
large-scale momentum and kinetic energy be negligible ---an inviscid criterion--- 
naturally supports a density weighted coarse-graining 
of the velocity field. Such a coarse-graining method is already known in the literature
as Favre filtering; however its use has been primarily motivated by appealing modeling
properties rather than underlying physical considerations. 
We also prove that  kinetic energy injection can be localized to the largest scales
by proper stirring, and argue that  stirring with an external acceleration field rather than a body force would yield a longer inertial range in simulations. 
We then discuss the special case of buoyancy-driven flows subject to a spatially-uniform gravitational field. 
We conclude that a range of scales can exist over which the mean kinetic 
energy budget is dominated by inertial processes and is immune from contributions
due to molecular viscosity and external stirring.
\end{abstract}
\end{center}

\vspace{0.5cm}

~~~{\bf Key Words:} Compressible turbulence, Scale decomposition, Inertial range, Favre filtering
\clearpage

\section{Introduction}

The aim of this paper is to give a systematic, theoretical approach
based on coarse-graining (or filtering) to analyze non-linear scale 
interactions in compressible turbulent flows. It builds upon previous work of Germano \cite{Germano92} 
and Eyink \cite{Eyink95a,Eyink05}.
There are several motivations for this work.
First, there is no unique way to specifying a notion of scale,
such as defining large-scale momentum and large-scale kinetic energy, in compressible turbulence.
The traditional approach in this subject has employed
density-weighted averaging, also known as Favre averaging, to decompose
a flow into large-scale and turbulent components \cite{Favre69}. 
Using Favre averaging, density, $\rho(\bx)$, and velocity, $\bu(\bx)$,
are combined to yield a large-scale momentum,
$\langle\rho\bu\rangle$, and a large-scale kinetic energy,
$\frac{1}{2}\langle\rho\bu\rangle^2/\langle\rho\rangle$, 
where $\langle\dots\rangle$ can denote an ensemble-average, a time-average, 
or, as used in this paper, a space-average $\frac{1}{V}\int_Vd\bx (\dots)$.
However, such a decomposition has been primarily borne out of convenience 
to modelers and practitioners rather than physical
considerations. It seems that {\it a priori} there is no fundamental 
reason to favor Favre averaging over any other combination of 
$\rho$ and $\bu$, from an infinite number of possibilities, such as defining large-scale kinetic energy
as $\langle\rho^\alpha \bu\rangle\bdot\langle\rho^{1-\alpha} \bu\rangle/2$ for $0<\alpha<1$ \cite{KidaOrszag90a} or, alternatively, as $\langle\rho\rangle|\langle \bu\rangle|^2/2$ \cite{Chassaing85}.

Second, and more importantly, this paper provides the foundation for recent work 
\cite{Aluie11,AluieLiLi12} in which we addressed basic questions about the nature of the cascade in
compressible turbulence. While the classical ideas of Richardson, Kolmogorov, and Onsager
form the cornerstone for our modern understanding of incompressible turbulence, there had been 
no grounds  for extending such a theory to compressible flows. The potent ideas of an inertial range 
and universality are often invoked without physical basis in  
compressible turbulence. Elementary questions on the possible existence of 
a scale-range which is immune from direct effects of viscosity and large
scale forcing, on whether energy is transferred to small scales through a 
cascade process, and whether such a cascade is local in scale, had not been 
previously addressed. Resolving these questions is necessary to warrant
the concept of an inertial range and to justify the existence of universal
statistics of turbulent fluctuations. Furthermore, Kolmogorov's 4/5-th law for the energy 
flux is an exact result which has no counterpart in compressible 
turbulence. An analogous result would be essential for attempting to 
predict the scaling of spectra and structure functions. Recent progress in this regard has been
encouraging; Falkovich {\it et al.} \cite{Falkovichetal10} derived a relation for compressible turbulence
analogous to the Karman-Howarth relation for incompressible turbulence,
and Galtier and Banerjee \cite{GaltierBanerjee11} derived an exact relation for the special case of
compressible turbulence with an isothermal equation of state.

This work is also very useful from the standpoint of numerical modeling.
Compressible flows, especially in astrophysical systems,
often involve a huge range of scales which cannot be simulated
directly. The coarse-graining approach provides
a theoretical basis for constructing models of
turbulence that may faithfully reflect the dynamics at unresolved scales.
The formalism that we employ is the same as that used in 
large-eddy simulation (LES) modeling of turbulent flows. This work thus 
provides a theoretical complement to those modeling efforts.
However, while the equations we analyze coincide (to a considerable extent) with 
those that are employed in LES of compressible turbulence,
their use here will be for rather different purposes. In LES, plausible but
uncontrolled closures are adopted for the subscale terms, whereas the aim here, as 
in \cite {Eyink95a,Eyink05,Aluie11,AluieLiLi12},
is to develop several exact estimates and some general physical understanding 
of these terms. Another difference is that LES generally takes the scale parameter
$\ell$ to be a fixed length of the order of the ``integral scale'' $L$. Our interest here
is rather to probe \emph{all scales in the flow}, including limits of small $\ell\ll L$.

We shall show below how a decomposition based on Favre filtering
(and averaging) comes out naturally from the physical requirement that viscous effects 
have a negligible role in the dynamics of large-scale momentum and large-scale 
kinetic energy. We call this the \emph{inviscid criterion}.
Using the coarse-graining approach, we will prove rigorously the
existence of an intermediate range of scales over which viscous dissipation
and external kinetic energy injection can be made to vanish.
Our decomposition also leads to two terms responsible for transferring kinetic 
energy across scales and constitute the so-called ``subgrid scale flux,''
which we shall discuss in detail.

This paper is intended for an audience with diverse backgrounds. For that reason, we have attempted to present the underlying physical ideas at an intuitive level in the main text while leaving mathematical details and rigorous proofs to appendices. The outline is as follows. 
In Section \ref{sec:Preliminaries} we present preliminary definitions and discussion. 
In Section \ref{sec:viscousrange} we show how viscous dynamics can be isolated to 
the smallest scales by a proper scale-decomposition and
in Section \ref{sec:inertialscalerange} we discuss the inertial dynamics based on such a decomposition.
In Section \ref{sec:Injection} we prove that it is possible to localize kinetic energy injection
to the largest scales in a system and discuss buoyancy-driven (Rayleigh-Taylor) flows 
as a special case.
In Section \ref{sec:Compressibilityeffects}, we examine contributions from compressibility effects to the 
flux of kinetic energy across scales. We conclude with Section \ref{sec:summary} and
three appendices which contain detailed proofs of our results.

\section{Preliminaries\lb{sec:Preliminaries}}

\subsection{Governing dynamics}
In this paper we study the dynamics at various scales through
a direct analysis of the compressible Navier Stokes equations without the use of
any closure approximation. The equations are those of continuity (\ref{continuity}), momentum (\ref{momentum}), 
and either internal energy (\ref{internal-energy}) or total energy (\ref{total-energy}):
\begin{eqnarray} 
&\hspace{-0.4cm}\partial_t \rho& + \partial_j(\rho u_j) = 0 \lb{continuity} \\
&\hspace{-0.4cm}\partial_t (\rho u_i)& + \partial_j(\rho u_i u_j) 
= -\partial_i P +  \partial_j\sigma_{ij} + \rho F_i   \lb{momentum}\\
&\hspace{-0.4cm}\partial_t (\rho e)&+ \partial_j\left\{\rho e u_j \right\}
= -P\partial_j u_j + 2\mu(|S_{ij}|^2 - \frac{1}{d}|S_{kk}|^2) -\partial_j q_j  ~~~~~~~~\lb{internal-energy}\\
&\hspace{-0.4cm}\partial_t (\rho E)&+ \partial_j(\rho E u_j) 
= -\partial_j (P u_j) +\partial_j[2\mu ~ u_i(S_{ij} - \frac{1}{d} S_{kk}\delta_{ij})]  -\partial_j q_j +\rho u_i F_i  \lb{total-energy}\\
&\hspace{-0.4cm}\partial_t (\rho \frac{|\bu|^2}{2})&\hspace{-0.4cm}+ \partial_j\left\{(\rho \frac{|\bu|^2}{2} + P) u_j -2\mu(u_i S_{ij} - \frac{1}{d}u_j S_{kk})\right\}
= P\partial_j u_j - 2\mu(|S_{ij}|^2 - \frac{1}{d}|S_{kk}|^2)  +\rho u_iF_i. \hspace{.8cm}\lb{kinetic-energy}\\
\nonumber\end{eqnarray}
We also write down the kinetic energy budget (\ref{kinetic-energy}) for convenience.
Here, $\bu$ is velocity, $\rho$ is density, $e$ is internal energy per unit mass, $E=|\bu|^2/2 + e$ is total energy per unit mass,  $P$ is pressure, 
$\mu$ is dynamic viscosity, ${\bf F}$ is an external acceleration field stirring the fluid, 
${\bf q} = -\kappa \grad T$ is the heat flux with a conduction coefficient $\kappa$ and temperature $T$. 
The symmetric strain tensor is $S_{ij} = (\partial_j u_i + \partial_i u_j)/2$, and the viscous stress, $\sigma_{ij}=2\mu(S_{ij} - d^{-1} S_{kk}\delta_{ij})$, is deviatoric (traceless) in $d$-dimensions. For convenience, we have assumed a zero bulk viscosity even though all our analysis applies to the more general case. Dynamic viscosity varies in space and is well-described by Sutherland's law or simpler power law approximations, $\mu(\bx)\sim T^\alpha(\bx)$.

\subsection{Coarse-graining\lb{sec:coarsegraining}}
We first present a general approach to analyzing scale interactions in a turbulent flow.
Following  Leonard \cite{Leonard74}, Germano \cite{Germano92}, and Eyink \cite{Eyink05}, we use a simple filtering technique common in the LES literature to resolve turbulent fields simultaneously in scale and
in space.

For any field $\ba(\bx)$,
a ``coarse-grained'' or (low-pass) filtered field, which contains modes
at scales $>\ell$, is defined in $d$-dimensions as
\be
\OL \ba_\ell(\bx) = \int d^d\br~ G_\ell(\br) \ba(\bx+\br),
\lb{filtering}\ee
where $G(\br)$ is a convolution kernel. It can be any real-valued function which is sufficiently smooth,
decays sufficiently rapidly for large $r$, and is normalized so that $\int d^d\bs ~G(\bs)=1$ for dimensionless $\bs$. It is assumed
furthermore that $G$  is centered, $\int d^d \bs ~\bs \,G(\bs) = \bzed$, and with the main support
in a ball of unit radius, $\int d^d \bs ~|\bs|^2 G(\bs) = \mathcal{O}(1)$. 
Its dilation in a $d$-dimensional domain $\Omega$, $G_\ell(\br)\equiv \ell^{-d} G(\br/\ell)$, will share these properties 
except that its main support will be in a ball of radius $\ell$. If $G(\bs)$ is also non-negative,
then (\ref{filtering}) may be interpreted as a local space average. Note that $G(\bs)$ can be
chosen so that	both it and its Fourier transform $\hat{G}(\bk)$ are positive and infinitely differentiable,
with $\hat{G}(\bk)$ also compactly supported inside a ball of radius $1$ about the origin in Fourier space
and with $G(\bs)$ decaying faster than any power $|\bs|^{-p}$ as $|\bs|\to\infty$. See for instance
Appendix A in \cite{EyinkAluie09} for explicit examples.
It can be shown that for any kernel $G(\bs)$ with the above properties, a coarse-grained function $\OL{f}_\ell(\bx)$
is infinitely differentiable 
\footnote{Under the very weak requirement that $\int_{\Omega}d\bx |f(\bx)|<\infty$ over the domain $\Omega$ of the flow.}.

We can also define a complementary high-pass filter which retains only modes
at scales $<\ell$ by
\be  \ba^{'}_\ell(\bx) = \ba(\bx)-\OL\ba_\ell(\bx).
\lb{high-pass}\ee
In the rest of our paper, we shall take the liberty of dropping subscript  $\ell$ whenever there is no risk of 
ambiguity.

It has been remarked by \cite{SagautGermano05} and \cite*{Garnieretal09} that
filtering a strong discontinuity in a field, such as an external shock from an explosion,
introduces ``parasitic'' contributions which can overwhelm the turbulent fluctuations at small scales.
Since our primary purpose in this work is a fundamental physical understanding rather than modeling
of non-linear scale interactions, and since such a strong shock would interact with the flow,
we consider it only natural to include its contributions to sub-scales $<\ell$.

The filtering operation (\ref{filtering}) is linear and commutes with space (and time) derivatives. 
We can apply it to the continuity and momentum equations (\ref{continuity})-(\ref{momentum}) to describe 
dynamics of large-scale fields. However, as we mentioned above, there is no unique way to filter these equations. For 
example, we may define a large-scale momentum field either as $\OL{\rho}_\ell \OL{\bu}_\ell$ or
as $\OL{\rho\bu}_\ell$. Similarly, a large-scale kinetic energy may be defined as $\frac{1}{2}\OL{\rho}_\ell |\OL{\bu}_\ell|^2$
or $\frac{1}{2}|\OL{\sqrt{\rho}\bu}_\ell|^2$.

\section{Identifying the viscous range\lb{sec:viscousrange}}

A key idea of this paper is that the scale-decomposition of momentum and kinetic energy should satisfy 
the \emph{inviscid criterion}, {\it i.e.} it  should guarantee that
viscous contributions are negligible at large enough length-scales. 
This is necessary for the study of inertial range dynamics if such a scale-range exists in compressible turbulence. 

\subsection{Scale decomposition\lb{sec:scaledecomposition}}
Assume for simplicity that $\mu(\bx)=\mu$ is independent of spatial position $\bx$ such as in the case of isothermal flows. We shall consider the more general case afterward. Coarse-graining eq.(\ref{momentum}) and commuting the filter with space derivatives in the viscous diffusion terms yields

\be \partial_t \OL{\rho u}_i +\partial_j (\OL{\rho u_i u_j} )
= -\partial_i\OL{P}  + \mu \partial_j \{  (\partial_j \OL{u}_i + \partial_i \OL{u}_j) - \frac{2}{d} \partial_k \OL{u}_k \,\delta_{ij} \} + \OL{\rho F}_i.
\lb{largemomentum}\ee
With such a decomposition, the functional form of viscous terms in (\ref{largemomentum}) is similar
to their counterpart in incompressible flows despite the additional contributions due to compressibility. 
If $u^2_{rms}= \int d\bx |\bu(\bx)|^2<\infty$,
it can be shown rigorously that each of the viscous terms in eq. (\ref{largemomentum}) is 
bounded by $\const \mu \,u_{rms}/\ell^{2} (L/\ell)^{3/2}$ at every point $\bx$. 
Therefore, the large-scale momentum, defined as $\OL{\rho\bu}_\ell$, does not diffuse
under the action of molecular viscosity when $\mu \,u_{rms}/\ell^{2} \ll 1$.
The type of proof used is standard in real analysis and
for applications in turbulence theory, see \cite{EyinkNotes} and \cite{AluieThesis}.
We detail the proof in \ref{ap:Viscouseffects}, Proposition \ref{Prop1}. The idea behind
it is simple and purely kinematic; a term $ \mu \nabla^2 \OL{\bu}$ involves derivatives of a smooth filtered field.
Therefore, such gradients cannot become arbitrarily large as $\mu \to 0$, even though unfiltered gradients, $\grad\bu$,
may become unbounded. In fact, $\mu \nabla^2 \OL{\bu}$ may be expressed\footnote{See remark following Proposition \ref{Prop1}} in terms of quantities at scale $\ell$, namely a big-$O$ bound $O(\mu \,\delta u(\ell)/\ell^2)$
which becomes negligible as $\mu\to 0$. Here, an increment is $\delta \bu(\boell)=u(\bx+ \boell) - u(\bx)$.

The reader might question the value of a careful proof when one can arrive 
at the same conclusion by a simple dimensional argument. 
To illustrate the potential pitfalls of dimensional reasoning here,
consider the quantity $\mu\OL{\grad\bu:\grad\bu}_{\ell}$.
It may be argued that this should also vanish as $\mu\to 0$ for some fixed $\ell>0$.
However, it is well known in turbulence literature that it does not 
(see for example \cite{Sreenivasan84, Sreenivasan98,Pearsonetal04}).
The problem lies in that $\mu\OL{\grad\bu:\grad\bu}_{\ell}$ cannot be rewritten as derivatives of filtered fields.
In other words, it cannot be expressed in terms of quantities at scale $\ell$,
such as $O(\mu \, \delta u(\ell)^2/\ell^2)$, as one might innocently expect.
When $\mu\to 0$ gradients can become unbounded 
and the term $\OL{\grad\bu:\grad\bu}_{\ell}$ diverges.  
Phrased in the language of Fourier analysis, even though 
$\OL{\grad\bu:\grad\bu}_{\ell}$ has small wavenumber modes $<K\sim\ell^{-1}$, 
it can be dominated by contributions from $\hat{\bu}(\bq)$ with wavenumbers 
$|\bq|\gg K$ due to the convolution $\grad\bu:\grad\bu$ in Fourier space. 
Here, 
$$\widehat{\bu}(\bq) = \int d\bx ~\bu(\bx) e^{-i \bq\bdot\bx}$$ 
is a Fourier transform. Put more explicitly, while the product $\widehat{\grad\bu}(\bq): \widehat{\grad\bu}(\bk-\bq)e^{i\bq\bdot\bx}e^{i(\bk-\bq)\bdot\bx}$
has a Fourier mode at wavenumber $|\bk|<K$, it is proportional to $\sim |\bq|^2$.
This example has a direct bearing on our definition of large-scale momentum.
If we were to define large-scale momentum as $\OL\rho_\ell\OL\bu_\ell$ rather than 
$\OL{\rho\bu}_\ell$ as we did above, a viscous term
in the balance equation would have the form $\mu \OL\rho_\ell \OL{\rho^{-1}\nabla^2 \bu}_\ell$.
Here, again, the filtering operation would not commute with the laplacian and,
due to possibly dominant contributions from high wavenumber modes $\gg \ell^{-1}$,
we would not be able to guarantee {\it a priori}\footnote{It is possible to prove that
$\mu \OL\rho_\ell \OL{\rho^{-1}\nabla^2 \bu}_\ell$ vanishes with $\mu\to 0$
under very restrictive assumptions on smoothness of the density field. Put loosely,
such assumptions would correspond to a density spectrum decaying faster than $k^{-3}$,
which is physically unrealistic in highly compressible flows.} that viscous terms are negligible at large $\ell$.

The arguments above and the proof in Proposition \ref{Prop1} presume that $\mu$ is a constant. In Proposition \ref{Prop3}, we extend the result to the more general case of a spatially variable viscosity, $\mu(\bx)$, under an additional assumption that $\grad\mu$ vanishes with $\mu\to 0$. In \ref{ap:Viscouseffects_variable}, we provide a physical justification for this assumption based on Sutherland's law or a power law relation, $\mu(\bx) \sim T^{\alpha}(\bx)$.
{\it A priori} tests of compressible turbulence simulations by 
\cite*{Vremanetal95}, \cite*{VremanThesis}, and \cite*{Martinetal00}  seem to suggest that, 
indeed, the additional non-linearity in viscous terms introduced by a spatially varying $\mu(\bx)$ is small.

The scale decompostion employed in the large-scale momentum balance (\ref{largemomentum})
is equivalent to traditional Favre filtering
(see for example \cite{Garnieretal09}), 
where a Favre filtered function is weighted by the density:
\be \wt{f}_\ell(\bx) \equiv \frac{\OL{\rho f}_\ell(\bx)}{\OL\rho_\ell(\bx)}. \lb{FavreDef}\ee
The operator $(\,\wt{\cdot}\,)$ is linear but does not commute with derivatives.
The large-scale momentum balance (\ref{largemomentum}) can be rewritten using 
definition (\ref{FavreDef}) as
\begin{eqnarray} 
\partial_t \OL\rho \wt{u}_i + \partial_j (\OL\rho \wt{u}_i~\wt{u}_j )
 =  -\partial_j\left(\OL\rho~\wt\tau(u_i,u_j)\right) -\partial_i\OL{P}  + \partial_j \OL{\sigma}_{ij} + \OL{\rho} \wt{F}_i.
\lb{Favremomentum}\end{eqnarray}
This is the same as the ``bare'' momentum equation (\ref{momentum}) itself but with an additional
contribution from \emph{turbulent stress},
\be\OL\rho\wt\tau(u_i,u_j)\equiv \OL\rho(\wt{u_iu_j} - \wt{u}_i~ \wt{u}_j),\lb{Favrestress}\ee
which accounts for the effect of eliminated scales $<\ell$ and vanishes identically 
in the absence of fluctuations at those small scales.
One can also obtain a continuity equation for large-scale density:
\be
\partial_t \OL{\rho} + \partial_i(\OL\rho \wt{u}_i) = 0.
\lb{Favrecontinuity}\ee

A main advantage of the filtering approach to analyzing turbulent flows is an ability to
resolve the relevant physical processes both \emph{in scale} and \emph{in space}
as is apparent from the balance eqs. (\ref{Favremomentum}),(\ref{Favrecontinuity}).
They describe the evolution of large-scale momentum and large-scale density
at every $\bx$ in the flow and at variable resolution $\ell$. Using $\OL\rho_\ell$ and $\OL{\rho\bu}_\ell$, and 
eqs. (\ref{Favremomentum}),(\ref{Favrecontinuity}), it is also straightforward to derive
a budget for kinetic energy density at scales $>\ell$, for arbitrary $\ell$.
This yields
\be
\partial_t \OL\rho_\ell\frac{|\wt\bu_\ell|^2}{2} + \grad\bdot\bJ_\ell
= -\Pi_\ell -\Lambda_\ell + \OL{P}_\ell\grad\bdot\OL\bu_\ell
-D_\ell
+\epsilon^{inj}_\ell,
\lb{largeKE}\ee
where $\bJ_\ell(\bx)$ is space transport of large-scale kinetic energy, $\Pi_\ell(\bx)+\Lambda_\ell(\bx)$,
which we examine closely in section \ref{sec:inertialscalerange}, is usually called
the \emph{subgrid scale (SGS) kinetic energy flux} to scales $<\ell$, $-\OL{P}_\ell\grad\bdot\OL\bu_\ell$ is large-scale
\emph{pressure dilatation}, $D_\ell(\bx)$ is viscous dissipation acting
on scales $>\ell$, and $\epsilon^{inj}_\ell(\bx)$ is the energy injected due to external stirring. These
terms are defined as 
\begin{eqnarray}
&\Pi_\ell(\bx)& = ~  -\OL{\rho}~ \partial_j\wt{u}_i  ~\wt\tau(u_i,u_j) ~~ \lb{flux1}\\[0.3cm]
& \Lambda_\ell(\bx)& = ~  \frac{1}{\OL\rho}\partial_j\OL{P}~\OL\tau(\rho,u_j) ~~ \lb{flux2}\\[0.3cm]
&D_\ell(\bx)&  =  \partial_j\wt{u}_i \left[2\,\OL{\mu\, S_{ij}} - \frac{2}{d}\, \OL{\mu\, S_{kk}}\,\delta_{ij} \right] \\[0.3cm]
&J_j(\bx)&  =  \OL\rho\frac{|\wt\bu|^2}{2}\wt{u}_j + \OL{P}\OL{u}_j + \wt{u}_i\OL\rho\wt\tau(u_i,u_j) 
-\wt{u}_i \OL{\sigma}_{ij}   \\[0.4cm]
&\epsilon^{inj}_\ell(\bx)& =  \wt{u}_i ~\OL{\rho} \wt{F}_i\lb{injectiondef}\\
\nonumber \end{eqnarray}
where we employed in (\ref{flux2}) the notation
\be  \OL\tau_\ell(f,g) \equiv\OL{(fg)_\ell}-\OL{f}_\ell\OL{g}_\ell \lb{tau-def} \ee
for $2^{nd}$-order \emph{generalized central moments} 
of any fields $f(\bx),g(\bx)$ (see \cite{Germano92}). 

Just as we have shown that viscous diffusion of large-scale momentum $\OL{\rho\bu} = \OL{\rho}\widetilde\bu$ is negligible, we can also rigorously prove under very weak conditions that viscous dissipation $D_\ell(\bx)$  of large-scale kinetic energy $\frac{1}{2}\OL\rho |\wt\bu|^2$ vanishes at every point $\bx$ when $\mu\, u^2_{rms}/\ell^2 \ll 1$. The rigorous proofs for both cases of a constant and spatially varying $\mu$ are given in Propositions \ref{Prop2} and \ref{Prop4} of \ref{ap:Viscouseffects}, respectively.

\subsection{Favre filtering and the inviscid criterion\lb{sec:physicalcriterion}}
In one of his original articles \cite{Favre69}, Favre  motivated the usage of density-weighted averaging by the 
fact that average mass of a fluid in a volume $V$ advected  by the 
large-scale velocity $\langle\rho\bu\rangle/\langle\rho\rangle$ is conserved\footnote{
There were other factors in Favre's choice of ``M\'ethode $\langle {\mathrm{B}}\rangle$,'' as he termed it, including a simpler form of the resultant equations, and applicability to experimental measurements (e.g. see \cite{GatskiBonnet09} for an overview).}.
We shall now briefly repeat Favre's argument. The change of average mass in a volume $V$ advected with some 
large-scale velocity $\bu^*$ is
\be \int_V d^3\bx ~\partial_t \langle\rho\rangle + \grad\bdot \left(\langle\rho\rangle\bu^* \right)
=\int_V d^3\bx ~ \grad\bdot \left(\langle\rho \rangle\bu^*  - \langle\rho\bu\rangle\right),
\ee
where now $\langle\dots\rangle$ denotes ensemble averaging. 
The equality follows from using the ensemble averaged
continuity eq. (\ref{continuity}). Favre averaged velocity $\wt\bu$ is defined as the choice of $\bu^*$ which makes
average flux of mass due to fluctuations (or turbulence) vanish, 
$\langle\rho \rangle\wt\bu  - \langle\rho\bu\rangle \coloneqq 0$. The same argument carries
over to spatially filtered dynamics, where now $\wt\bu_\ell$ is defined as the choice of large-scale
velocity which does not lead to subgrid terms in the filtered continuity eq. (\ref{Favrecontinuity}).

There are two comments we would like to make concerning Favre's argument. 
First, the special property large-scale velocity $\wt\bu$ enjoys, {\it i.e.} suppressing turbulent diffusion of mass 
in eq. (\ref{Favrecontinuity}), does not logically imply by itself that a density-weighted decomposition
is a necessary choice in the mass balance (\ref{Favrecontinuity}).  It is certainly not unphysical for 
a turbulent flow to diffuse mass and, in this respect, the choice of a large-scale velocity field
would depend on the particular purpose of an investigation. 
Second, a criterion requiring that turbulent \emph{mass} diffusion be zero has no logical implication
on the scale decomposition of  \emph{momentum} and \emph{kinetic energy}. Unlike in eq. (\ref{Favrecontinuity}), 
the Favre decomposition results in turbulent diffusion and dissipation of large-scale momentum and 
kinetic energy as seen from eqs. (\ref{Favremomentum}),(\ref{largeKE}).

Yet, density-weighted filtering is used extensively in LES of compressible turbulence 
due to its modeling appeal. One of the perceived advantages is the absence of subgrid 
terms to be modeled in the coarse-grained continuity eq. (\ref{Favrecontinuity}).
Another reason is that Favre filtered equations (\ref{Favremomentum}),(\ref{largeKE}) 
are structurally similar to their classically filtered counterparts in incompressible flows,
which allows for carrying over models from the incompressible LES literature.
Furthermore, none of the subgrid scale terms 
is a function of pressure which practitioners try to avoid modeling. When using the ideal gas law,
$P=\const \rho \,T$, as the equation of state, there is also an added advantage that filtered pressure, $\OL{P}$,
can be expressed as a function of resolved quantities, $\OL\rho$ and $\wt{T}$, without additional subgrid
terms.

What we have shown above is that Favre decomposition of momentum and kinetic energy
satisfies the \emph{inviscid criterion}. It guarantees that viscous contributions are negligible 
at large enough length-scales. Such a decomposition of 
momentum and kinetic energy is borne out of a physical requirement, irrespective 
of practical modeling considerations. We remark, however, that it may not be
the unique decomposition satisfying the inviscid criterion. In other words, we did not prove that 
it is necessary. We only showed that the Favre decomposition is sufficient to satisfy the 
inviscid criterion.

As we mentioned in the introduction,
while our equations (\ref{Favremomentum}),(\ref{Favrecontinuity}),(\ref{largeKE}) 
coincide to a considerable extent with those that are employed in LES of 
compressible turbulence, their use here and in the ensuing papers will be for rather different purposes. 
Whereas the primary goal in LES is to model the subgrid terms, the aim here
is to develop a physical understanding of these terms and estimate their
contributions at different scales, including limits of small $\ell\ll L$,
 through exact mathematical analysis.

\section{Inertial dynamics \lb{sec:inertialscalerange}}
Now that we have isolated viscous effects to the smallest scales $\ell_\mu$,
where $\ell_\mu$ is defined as the scale at which viscous effects become significant
in kinetic energy balance (\ref{largeKE}), we can study the dynamics at scales $\ell\gg\ell_\mu$.

\subsection{Deformation work}
The first term in kinetic energy SGS flux, $\Pi_\ell$ in eq. (\ref{largeKE}), is the contribution from 
\emph{deformation work} done by \emph{large-scale strain}
$\partial_j\wt{u}_i$ against the \emph{subgrid stress} $\OL\rho\,\wt\tau(u_i,u_j)$ (see for example \cite{TL}).
This is similar to its incompressible counterpart except that the strain is not traceless
here. It acts as a sink in the large-scale kinetic energy budget (\ref{largeKE}) and 
as source in the complementary small-scale kinetic budget (\ref{smallKE}),
and represents that part of the kinetic energy transferred from
scales larger than $\ell$ to smaller scales at point $\bx$ in the flow. 

Furthermore, $\Pi_\ell(\bx)$ is Galilean invariant due to the subtracted large-scale terms in definition
(\ref{Favrestress}) of the turbulent stress. Other definitions of the SGS flux are possible
such as $\wt{u}_i \partial_j(\OL\rho\wt\tau(u_i,u_j))$ which differs from our definition
(\ref{flux1}) by a total gradient $\partial_j(\OL\rho \wt u_i \wt\tau(u_i,u_j))$. However,
this definition is not pointwise Galilean invariant, so the amount of ``energy cascade'' at any point
$\bx$ in the fluid according to such a definition would depend on the observer's velocity.
Kraichnan \cite{Kraichnan64},
Speziale \cite{Speziale85}, and Germano \cite{Germano92} all emphasized the importance of Galilean invariance.
More recently, Eyink and Aluie \cite{EyinkAluie09,AluieEyink09,AluieEyink10} showed that Galilean invariance was necessary for scale-locality of the cascade.
There are non-Galilean-invariant terms in our budget (\ref{largeKE}) but, as is physically natural,
they are all associated with space transport $\bJ$ of kinetic energy.

Another physical requirement on the flux $\Pi_\ell(\bx)$ is that it should vanish
in the absence of fluctuations at scales smaller than $\ell$ (or a moderate fraction thereof); 
for example, when $\ell$ is 
equal to $2\pi/K_{max}$, where $K_{max}$ is the maximum wavenumber in a numerical simulation 
\cite{AluieKurien11}. 
This is satisfied by our definition of $\Pi_\ell(\bx)$ identically at every point $\bx$ in the flow. 
Other definitions of an energy flux are possible, such as the ``unsubtracted flux'' of an incompressible flow
critiqued in \cite{EyinkAluie09,AluieEyink09},
$$\Pi_\ell^{uns}(\bx)\equiv \OL{u}_i ~\OL{\cal NL}_{i}, \,\,\,\,\,\,\,\,\,{\cal NL}_{i}=\partial_j(\rho u_i u_j)
$$
which is often employed in literature that considers the sharp-spectral filter. 
Here, ${\cal NL}_{i}$ denotes the nonlinearity in the momentum equation.

Using such a filter, the ``unsubtracted flux'' across wavenumber $K$ is computed as 
$$\Pi_K^{uns}(\bx) = \sum_{|\bp|<K}\hat{u}_i(\bp)e^{i\bp\cdot\bx} \sum_{|\bk|<K}\widehat{{\cal NL}_{i}}(\bk) e^{i\bk\cdot\bx}.
$$
Taking $K = K_{max}$, we have $\Pi_{K_{max}}^{uns}(\bx) = u_i \partial_j(\rho u_i u_j)$ which is in general nonzero.
It is only after averaging over all space (and in the absence of flow beyond the domain boundary) that one gets
$$\langle\Pi_{K_{max}}^{uns} \rangle =\langle\partial_j(\rho\frac{|\bu|^2}{2}u_j ) \rangle= \sum_{|\bk|<K_{max}} \hat{u}^{*}_i(\bk) ~\widehat{{\cal NL}_{i}}(\bk)  = 0.
$$

Similar considerations apply for compressible flows, where an unsubtracted flux may be defined as 
$$\Pi_\ell^{uns}(\bx)\equiv \wt{u}_i ~\OL{\cal NL}_{i} - \frac{1}{2}|\wt\bu|^2~\OL{\cal N}
, \,\,\,\,\,\,\,\,\,{\cal NL}_{i}=\partial_j(\rho u_i u_j), \,\,\,\,\,\,\,\,\,{\cal N}=\partial_j(\rho u_j),
$$
with ${\cal NL}_i$ and ${\cal N}$ denoting the nonlinearities in momentum and density
equations, respectively. This flux does not vanish in general when $\ell=K^{-1}_{max}$, except after space-averaging.

\subsection{Baropycnal work}
The other part of kinetic energy flux, $\Lambda_\ell$ in eq. (\ref{largeKE}), is intrinsically due
to compressibility effects and vanishes in the absence of density variations. 
It represents work done by a \emph{large-scale pressure-gradient force}\footnote
{The term``pressure-gradient force'' is used in the meteorology literature.
It is not a force but an acceleration.}
$\OL{\rho}^{-1}\grad\OL{P}$
against \emph{subscale mass flux}\footnote{Here, ``flux'' denotes a flux in space, not in scale.} 
$\OL{\tau}(\rho,\bu)$. 
We shall refer to $\Lambda_\ell(\bx)$ as \emph{baropycnal work} due to its 
inherent dependence on pressure and density variations. It is not entirely due to 
baroclinic effects for it can be non-zero even when small-scale density variations and 
$\grad\OL{P}_\ell$ are aligned as shown in Figure \ref{Fig:Lambda}.
Similar to $\Pi_\ell$, it also acts as a sink in the large-scale
kinetic energy budget (\ref{largeKE}) and as a source in the small-scale budget (\ref{smallKE}), is pointwise Galilean invariant, and vanishes identically at every $\bx$
in the absence of fluctuations at scales $<\ell$. 
Baropycnal work is known to play a major role in turbulent combustion (see for example 
\cite{StarnerBilger80,LibbyBray81}).
It has also been recently observed in \cite{LivescuRistorcelli07,Livescuetal09}
to play a major role in turbulence production
in buoyancy-driven flows with significant density differences. In such flows,
termed ``variable-density flows'' in \cite{LivescuRistorcelli07},
there are two or more incompressible miscible fluid species, such as water and brine, which
have significantly different densities.
 \begin{figure}
\centering
\includegraphics[angle=0,totalheight=.3\textheight,width=.55\textwidth]{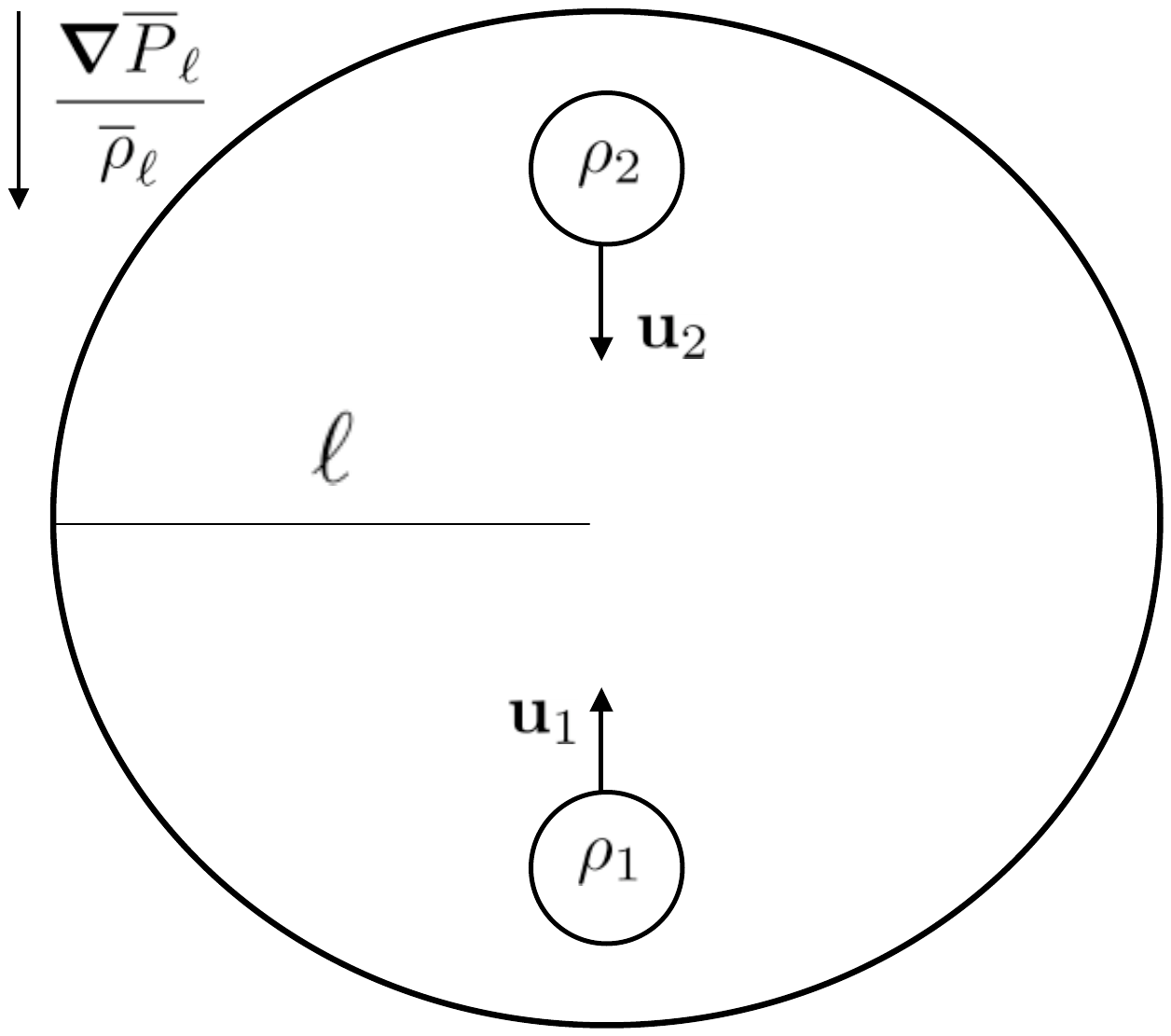}
\caption{Heuristic explanation of the physics behind baropycnal work, $\Lambda_\ell$.
In a ball of radius $\ell$, there are two small-scale fluid parcels of equal size 
with densities $\rho_1$ and $\rho_2$, and antiparallel velocities $\bu_1$ and $\bu_2$, respectively.
In a frame co-moving with the ball, we must have $|\bu_1|=|\bu_2|$.}
\lb{Fig:Lambda}\end{figure}

The physical mechanism 
behind this part of the SGS flux, illustrated in Figure \ref{Fig:Lambda}, is simple. 
In a frame \emph{co-moving} with a ball of radius $\ell$ in the flow,
a pressure-gradient force $\OL\rho^{-1}\grad\OL{P}_\ell$
at scales $>\ell$ acts on the ball  within which the fluid has non-uniform density (density variations at scales $<\ell$).
Per unit time, 
if $\rho_2 > \rho_1$ in Figure \ref{Fig:Lambda} and $|\bu_1|=|\bu_2|=u$, then parcel $2$ gains $\OL\rho^{-1}|\grad\OL{P}_\ell| \rho_2 u$ 
in kinetic energy from the large-scales and parcel $1$ loses $\OL\rho^{-1}|\grad\OL{P}_\ell| \rho_1 u$ to the large-scales.
On aggregate, scales $<\ell$ in this ball would gain kinetic energy from the large-scales at rate,
$\OL\rho^{-1}|\grad\OL{P}_\ell| (\rho_2-\rho_1) u$. This effect would vanish in the absence of density variations. Note that if
$|\bu_1|$ does not equal $|\bu_2|$, then the whole ball will have an average large-scale velocity 
$\OL\bu_\ell = \bu_1+\bu_2$ which does not play a role in such a process of inter-scale energy transfer
---hence the relevance of the premise of a co-moving frame.

Similar to deformation work, $\Pi_\ell(\bx)$, baropycnal work, $\Lambda_\ell(\bx)$, is not sign-definite.  
We expect that at points $\bx$ in the flow where large-scale pressure gradient $\grad\OL{P}_\ell(\bx)$
opposes the density gradient due to small-scale fluctuations ($\rho_2 >\rho_1$ in Figure \ref{Fig:Lambda}),
then a motion similar to that of a Rayleigh-Taylor instability would ensue such that velocities $\bu_1$ and
$\bu_2$ will be as illustrated in Figure \ref{Fig:Lambda}. We, therefore, expect that at such points in the
flow $\Lambda_\ell(\bx)$ will be positive and transfer energy to small-scales.
On the other hand, in regions such as shocks where large-scale pressure gradient is in the same direction as
the density gradient due to small-scale fluctuations ($\rho_2 <\rho_1$ in Figure \ref{Fig:Lambda}),
we expect that $\Lambda_\ell(\bx)$ will be negative and transfer energy to large-scales.
Whether, on average, the subscale mass flux $\OL\tau(\rho,\bu)$ would correlate positively or negatively 
with large-scale acceleration field $\OL\rho^{-1}\grad\OL{P}$ can be determined empirically.

Despite the recognition of $\Lambda_\ell$'s importance in turbulent combustion and 
variable-density flows, to the best of our knowledge, this term has never been studied
as a function of scale $\ell$. The presence of baropycnal work 
as a cascade mechanism to transfer kinetic energy between scales has not been appreciated in the literature.
Possibly due to Favre's original formulation \cite{Favre69} (see also a review by Lele \cite{Lele94}),
this term has often been lumped with $\OL{P}_\ell\grad\bdot\OL{\bu}_\ell$
in the form $\OL{P}_\ell\grad\bdot\wt{\bu}_\ell$ (plus an additional space-transport term) and treated as a 
large-scale pressure dilatation which does not require modeling. An exception is the work by
Huang {\it et al.} \cite{Huangetal95} who, from a modeling standpoint, 
espoused separating the two terms ($\Lambda_\ell$ and $\OL{P}_\ell\grad\bdot\OL{\bu}_\ell$) as we have 
done in eq. (\ref{largeKE}) on the belief that a density-weighted
decomposition should only be applied to the convective terms. In contrast to \cite{Huangetal95}, our reason
for keeping baropycnal work separate from 
pressure dilatation is due to a fundamental distinction between the two. Notice that 
both deformation work, $\Pi_\ell$, and baropycnal work, $\Lambda_\ell$, involve \emph{large-scale}
fields acting against \emph{small-scale} fluctuations. This makes them capable of transferring 
energy \emph{across} scales. On the other hand, large-scale pressure dilatation, $\OL{P}_\ell\grad\bdot\OL{\bu}_\ell$,
involves only large-scales and cannot transfer energy directly across scales.

\subsection{Pressure dilatation}
Large-scale \emph{pressure dilatation} $-\OL{P}\grad\bdot\OL{\bu}$ in eq. (\ref{largeKE})
represents conversion of large-scale kinetic energy to internal energy through compression. 
In the incompressible limit, this vanishes at every point $\bx$. Unlike $\Pi_\ell$ and $\Lambda_\ell$,
pressure dilatation only contains scales $>\ell$ (at least for filter kernels $\hat{G}$(\bk) compact in Fourier space).
Therefore, it is not involved in the transfer of energy \emph{across} scale $\ell$ and it does not
vanish in the absence of subscale fluctuations. We recently analyzed this term using data from 
high-resolution numerical simulations \cite{AluieLiLi12} and concluded that pressure dilatation
acts primarily at large-scales, on average.

\section{Kinetic energy injection\lb{sec:Injection}}
Similar to localizing viscous dissipation to the smallest scales in a flow,
localizing kinetic energy injection to the largest scales is just as important to enable the study of an
intermediate scale-range over which inertial processes dominate.

When density is constant, $\rho=\rho_0$, both kinetic energy, $\rho_0 |\bu|^2/2$, and its injection, $\epsilon^{inj}=\rho_0\bu\bdot{\bf F}$, are quadratic quantities. This allows kinetic energy injection to be easily localized to the largest scales by restricting ${\bf F}$ to small Fourier wavenumbers as is commonly done. To recap why that this, we will denote a field in a $d$-dimensional periodic domain $\mathbb{T}^d=[0,2\pi)^d$, coarse-grained with 
the sharp-spectral filter to retain only Fourier modes $|\bk|<K$, by 
\be\ba^{<K}(\bx) \equiv \sum_{|\bk|\le K}\hat{\ba}(\bk) e^{i\bk\cdot\bx}.
\lb{sharp-spectral-def}\ee 
This is similar to $\OL\ba_\ell(\bx)$ with $\ell$ being of the same order as $K^{-1}$.
Now consider an imposed acceleration ${\bf F}^{<K_0}$ with modes restricted to small wavenumbers $\le K_0$.
In an incompressible flow, mean ``large-scale'' kinetic energy at modes $\le K$ is $\rho_0|\bu^{<K}|^2/2$, and the mean injection into those modes reduces to
\be\rho_0\langle\bu^{<K} \bdot{\bf F}^{<K_0}\rangle = \rho_0\langle\bu^{<K_0} \bdot{\bf F}^{<K_0}\rangle,
\lb{incomp-injection}\ee
due to orthogonality of Fourier modes. Relation (\ref{incomp-injection}) shows that mean injection into modes $<K$ becomes independent of $K$ for $K\ge K_0$, implying that there is no net energy deposited by the forcing at modes $\ge K_0$. In other words, relation (\ref{incomp-injection}) shows kinetic energy is injected at wavenumbers $\le K_0$ when ${\bf F}$ is restricted to modes $\le K_0$. In the more general case of variable density flows, relation (\ref{incomp-injection}) no longer holds and
localizing the injection, now a cubic quantity, is not as obvious.

While most numerical simulations of forced compressible turbulence stir the flow with a large-scale 
acceleration of the form ${\bf F}^{<K_0}$ 
(e.g. \cite{KidaOrszag90a,Schmidtetal09,PetersenLivescu10,Federrathetal10,Kritsuketal07}), 
to the best of our knowledge, the first study to pose the issue of whether such stirring
restricts kinetic energy injection to the largest scales was by
Wagner {\it et al.} \cite{Wagneretal12}. However, the authors of \cite{Wagneretal12} 
did not investigate injection of kinetic energy but that of $\langle|\rho\bu|^2\rangle/2$.

We shall now show analytically that stirring with an acceleration field of the form
${\bf F}^{<K_0}(\bx)$ restricts kinetic energy injection to the largest scales (within the 
Favre scale-decomposition framework utilized in budget (\ref{largeKE}), 
where ``large-scale'' for kinetic energy is $\OL{\rho}|\wt{\bu}|^2/2$). 
We defer the rigorous proof to \ref{ap:Kineticenergyinjection}, but the logic behind it is simple and goes as follows.
\begin{eqnarray}
\langle \epsilon^{inj}_\ell \rangle 
&=& \left\langle \frac{\OL{\rho\bu}_\ell}{\OL\rho_\ell} \bdot \OL{\rho{\bf F}}_\ell \right\rangle \nonumber\\
&=& \left\langle \frac{\OL{\rho\bu}_\ell}{\OL\rho_\ell} \bdot \OL{\rho}_\ell\OL{{\bf F}}_\ell \right\rangle
+ \left\langle \frac{\OL{\rho\bu}_\ell}{\OL\rho_\ell} \bdot \OL\tau_\ell(\rho,{\bf F}) \right\rangle,
\end{eqnarray}
which is exact. When the forcing ${\bf F}(\bx) = {\bf F}^{<K_0}(\bx)$ varies at very large scales $L\sim K_0^{-1}\gg \ell$, we have $\OL{{\bf F}}_\ell(\bx)\approx \OL{{\bf F}}_L(\bx)$. Moreover, $\OL\tau_\ell(\rho,{\bf F})$, representing sub-scale fluctuations in $\rho$ and ${\bf F}$ at scales smaller than $\ell\ll L$, is negligible primarily because ${\bf F}(\bx)$ has no fluctuations at small scales. It follows that 
\begin{eqnarray}
\langle \epsilon^{inj}_\ell \rangle 
\approx  \left\langle \OL{\rho\bu}_\ell \bdot \OL{{\bf F}}_L \right\rangle 
\approx \left\langle (\rho\bu)^{<K_0} \bdot {\bf F}^{<K_0} \right\rangle=\langle \epsilon^{inj}_{K_0} \rangle,
\lb{localizedinjection}\end{eqnarray}
where the third expression follows from the orthogonality of Fourier modes. In \ref{ap:Kineticenergyinjection}, we present a rigorous proof of statement (\ref{localizedinjection}) under precise (and weak) conditions. Relation (\ref{localizedinjection}) shows that mean injection into scales $>\ell$ becomes independent of $\ell$ for $\ell\ll K^{-1}_0$. Result (\ref{localizedinjection}) also implies that injection at scales smaller than $\ell$, $\langle\epsilon^{\mbox{\tiny{small}}}_\ell\rangle$ in eq. (\ref{smallinjectiondef}), vanishes as $\ell K_0 \to 0$.  This proves the non-trivial possibility to make injection localized to the largest scales in variable density flows by employing an external acceleration field limited to wavenumbers $\le K_0$. Finally, we note that the acceleration field ${\bf F}^{<K_0}$ in momentum eq. (\ref{momentum}) is of a general form and can have both solenoidal and irrotational components.

In arriving at eq. (\ref{localizedinjection}), we had to stir with an \emph{external acceleration} field such that the force in (\ref{momentum}) is weighted by instantaneous density $\rho(\bx,t)$. Had we stirred the momentum equation with an \emph{external force} ${\boldsymbol {\mathcal F}}^{<K_0}$ instead of $\rho{\bf F}^{<K_0}$, the injection would have had the form $\langle \epsilon^{inj}_\ell \rangle = \big\langle \wt{\bu}_\ell\bdot{\boldsymbol {\mathcal F}}^{<K_0} \big\rangle$ In \ref{ap:Kineticenergyinjection}, Proposition \ref{Prop6}, we prove rigorously that it converges to $\big\langle \bu^{<K_0}\bdot {\boldsymbol {\mathcal F}}^{<K_0} \big\rangle$, for $\ell K_0\to 0$. However, the rate of convergence according to our bound (\ref{injectionconv_force}) is slower than that in (\ref{injectionconv}) obtained by stirring with an acceleration field. This suggests that stirring with an acceleration field would yield a longer intertial range in numerical simulations with significant density fluctuations, compared to stirring with a force.

\subsection{The inertial range}
Following our result (\ref{localizedinjection}) and assuming statistically steady-state conditions, the 
spatially averaged large-scale kinetic energy budget (\ref{largeKE}) at scales $K_0^{-1}\gg\ell\gg\ell_\mu$ becomes
\be
\langle\Pi_\ell\rangle + \langle\Lambda_\ell\rangle -\langle\OL{P}_\ell\grad\bdot\OL{\bu}_\ell\rangle= \langle \epsilon^{inj}_{K_0} \rangle= \const,
\lb{steadyKE}\ee
such that the sum of kinetic energy flux and pressure dilatation on the left hand side
is constant, independent of scale $\ell=K^{-1}$. In relation (\ref{steadyKE}), we assumed that 
none of the kinetic energy is transported beyond the domain boundary.
We also dropped viscous dissipation, $\langle D_\ell\rangle$, which we have proved to be negligible.

Relation (\ref{steadyKE}) in itself is not an analogue to Kolmogorov's $4/5$-th law since it contains the pressure dilatation term which does not involve energy transfer across scales. 
However, we have recently presented a sufficient condition along with a physical argument in \cite{Aluie11} which imply that $\langle\OL{P}_\ell\grad\bdot\OL{\bu}_\ell\rangle$ acts primarily on the largest scales, similar to $\langle \epsilon^{inj}_{K_0} \rangle$. We have also shown evidence from numerical simulations in \cite{AluieLiLi12} in support of this, {\it i.e.}
\be\langle\OL{P}_\ell\grad\bdot\OL{\bu}_\ell\rangle \approx \langle\OL{P}_{L_c}\grad\bdot\OL{\bu}_{L_c}\rangle,
\lb{largepressuredilatation}\ee
where $L_c$ is a large ``conversion'' length-scale similar to an integral scale. It follows that most of the net conversion between kinetic and internal energy takes place at scales larger than $L_c$ and equation (\ref{steadyKE}) becomes
\be
\langle\Pi_\ell\rangle + \langle\Lambda_\ell\rangle = \langle\OL{P}_{L_c}\grad\bdot\OL{\bu}_{L_c}\rangle + \langle \epsilon^{inj}_{K_0} \rangle = \const
\lb{steadyKE_II}\ee
over the scale-range $\ell_\mu \ll \ell \ll L_c$. A scale-independent  kinetic energy flux in eq. (\ref{steadyKE_II}) implies that kinetic energy cascades conservatively in a manner similar to energy cascade in incompressible turbulence. We presented evidence of such a cascade in \cite{AluieLiLi12}. Relation (\ref{steadyKE_II}) can be regarded as an analogue to Kolmogorov's $4/5$-th law for compressible turbulence.

\subsection{Special case: Rayleigh-Taylor flow\lb{sec:RTflow}}

We shall apply our result on localized injection to the case of a Rayleigh-Taylor 
flow driven by gravitational forces. We ask the following question: At what scales
does mean conversion of gravitational potential energy  to kinetic energy take place?

We shall show that, in the presence of a spatially uniform gravitational field ${\bf g}(\bx)={\bf g}$,
such conversion into kinetic energy only takes place at the largest scale ---that of the domain size. 
Consider the large-scale kinetic energy equation (\ref{largeKE}) where 
${\bf F}(\bx)$ is replaced with ${\bf g}$ in $\epsilon^{inj}_\ell$. We have from (\ref{injectiondef}) that net input 
of kinetic energy due to gravity is
\be\langle\epsilon^{inj}_\ell\rangle = \langle\wt\bu_\ell \bdot\OL{\rho}_\ell\wt{\bf g}_\ell\rangle 
 = \langle \OL{\rho\bu}_\ell \bdot{\bf g}\rangle = {\bf g}\bdot\langle\rho\bu\rangle
\lb{RTinjection}\ee
where the last equality follows from $\int d\bx \OL{f}_\ell(\bx) = \int d\br G_\ell(\br) \int d\bx f(\bx+\br) = \langle f\rangle$.
Result (\ref{RTinjection}) shows that mean injection is independent of scale $\ell$ and only takes place at the
scale of the domain, $L_{\mbox{\tiny{dom}}}$.

Put in more detail, the mean kinetic energy $\langle\OL{\rho}_\ell |\wt{\bu}_\ell|^2\rangle/2$ 
at scales $>\ell$ increases at a rate $\langle\wt\bu_\ell \bdot \OL\rho_\ell \wt{\bf g}_\ell \rangle$
due to gravitational forces. Consider a sequence of scales $\ell_1 > \ell_2 > \dots >\ell_n >\dots$
and the average rate of potential energy being converted into mean kinetic energy,
$\langle\epsilon^{inj}_{\ell_n}\rangle\equiv\langle\wt\bu_{\ell_n} \bdot \OL\rho_{\ell_n} \wt{\bf g}_{\ell_n} \rangle$,
at successively larger sets of scales $[\ell_n,L_{\mbox{\tiny{dom}}})$.
In general, as $n\to\infty$, $\langle\epsilon^{inj}_{\ell_n}\rangle$ approaches the total rate of potential energy 
converted into kinetic form. The fact that $\langle\epsilon^{inj}_{\ell_n}\rangle$ is independent of $\ell_n$ over
the entire scale-range $[0,L_{\mbox{\tiny{dom}}})$ implies that all conversion takes place at the domain scale.

Our argument demonstrates the power of the filtering approach\footnote{
In this particular case, with a constant ${\bf g}$ containing only a $k=0$ mode, the same conclusion would 
have also been possible by examining the injection, $\langle\epsilon^{\mbox{\tiny{small}}}_\ell\rangle$ in eq. (\ref{smallinjectiondef}), into mean subscale kinetic energy, $\langle\OL{\rho|\bu|^2}_\ell - \OL{\rho}_\ell |\wt{\bu}_\ell|^2 \rangle/2$, which may be easily determined to equal zero at any $\ell$. See \ref{ap:SmallKE}.}
in analyzing non-linear scale interactions.
We were able to arrive at our answer precisely because the filtering technique allows for probing a 
\emph{continuous range of scales}, in contrast to the traditional averaging approach.

The conclusion is probably non-intuitive at face value because, in a Rayleigh-Taylor flow,
``fingers'' of heavy fluid \emph{at any scale} penetrate the lighter fluid as
they descend, converting potential to kinetic energy. It seems
to contradict our result that conversion only takes place at the domain scale.
Key to understanding such an ostensible paradox
is the definition of kinetic energy based on a Favre scale-decomposition.
Mean kinetic energy at scale  $\ell$ may be rewritten as
$\langle|\OL{\rho\bu}_\ell|^2/\OL\rho_\ell\rangle/2$, which emphasizes
the central role of \emph{momentum} in defining scale. Indeed, result
(\ref{RTinjection}) demonstrates that it is only the $k=0$ mode of momentum
which participates in converting potential energy into kinetic energy at
scale with mode $k=0$, $\langle\rho\bu\rangle ^2/\langle\rho\rangle/2$.
While momentum at scale $\ell$ can have contributions from density and velocity at 
\emph{all} scales ---for example, mean momentum $\langle\rho\bu\rangle = \sum_{\bk}\hat{\rho}(\bk)\hat{\bu}(-\bk)$ 
--- the scale $\ell$ of kinetic energy depends on that of momentum $\OL{\rho\bu}_\ell$ and not
on, for instance, $\OL{\rho}_\ell\OL\bu_\ell$.

\section{Compressibility effects in the SGS flux\lb{sec:Compressibilityeffects}}
Our scale-decomposition allowed us to identify two SGS sinks for the large-scale kinetic energy budget,
namely deformation work, $\Pi_\ell$, and baropycnal work, $\Lambda_\ell$.

While $\Pi_\ell$ also represents a cascade mechanism in incompressible turbulence,
$\Lambda_\ell$ emerges from density fluctuations and thus is intrinsic to flows with variable 
density.
However, even deformation work has contributions from compressibility effects.
We have shown in \cite{Aluie11,Aluie11arxiv_II} that 
$\Pi_\ell = \OL\rho \partial_j\OL{u}_i\OL\tau(u_i,u_j) + \dots\mbox{8 terms}\dots $, using exact identities.
If we consider that part of $\Pi_\ell$ equal to $\OL\rho \partial_j\OL{u}_i\OL\tau(u_i,u_j)$,
we see that compressive modes with $\bk\bdot\hat{\bu}(\bk)\ne0$ can play an important role in
deformation work.
This is because the large-scale strain $\grad\OL\bu$ is not traceless.
The effect of compressive modes is best illustrated by the one-dimensional Burger's flow, 
\be \partial_t u + \partial_x (\frac{u^2}{2}) =  \nu \partial_{xx}u.
\lb{Burger}\ee
A large-scale kinetic energy budget analogous to (\ref{largeKE}) can be derived (see for example \cite{EyinkNotes}):
\be \partial_t \frac{|\OL{u}|^2}{2} 
+ \partial_x \bigg\{\frac{\OL{u}^3}{3} + \OL{u}~\frac{\OL\tau(u,u)}{2} -\nu\partial_x\frac{|\OL{u}|^2}{2}\bigg\} 
= -\Pi^{burg}- \nu |\partial_{x}\OL{u}|^2,
\ee
in which viscous dissipation on the right-hand-side is negligible and
\be\Pi^{burg}\equiv-\partial_x\OL{u}~\frac{\OL\tau(u,u)}{2}\ee 
is the only sink. There are no shearing motions in flow (\ref{Burger}). The only way energy
cascades to small scales is through $\Pi^{burg}$ which is solely due to compressive modes.
As is well-known, this cascade is manifested in the formation of shocks.

In general, a simple measure that quantifies kinematic role of compressibility on
deformation work (\ref{flux1}) is
\be\Pi_\ell^{comp}\equiv \Pi_\ell + \OL\rho\,\,\partial_j\OL{u^s}_i\,\,\OL\tau(u^s_i,u^s_j),
\ee
where the velocity $\bu=\bu^s+\bu^c$ is decomposed into solenoidal and irrotational components, $\bu^s$ and $\bu^c$,
respectively.

\section{Summary\lb{sec:summary}}

In this paper we have shown that viscous diffusion and dissipation can be isolated to 
the smallest scales in a compressible flow by using a proper scale-decomposition. Guided
by this physical requirement, which we call the ``inviscid criterion'', we found that a Favre 
decomposition of the momentum and 
kinetic energy into large-scale and small-scale components is sufficient to guarantee 
a negligible role of molecular viscosity in the large-scale dynamics of high Reynolds 
number flows.

We were also able to establish through an exact analysis that mean kinetic energy injection
can be made localized to the largest scales in a flow by proper stirring. Moreover, our analysis suggested
that stirring with an external acceleration field is more adequate to realizing a longer inertial range in a numerical simulation compared to stirring with an external force. We discussed the special case of 
buoyancy-driven flows in which stirring is due to a spatially-uniform gravitational field, 
and showed that mean injection of kinetic energy occurs only at the very largest scale in the 
system.

Localizing viscous dissipation to the smallest scales and energy injection to the largest
scales is necessary to allow for studying inertial dynamics at intermediate scales.
Under steady-state conditions, satisfying these two ingredients implies that the sum
of mean SGS kinetic energy flux and pressure dilatation, 
$\langle\Pi_\ell+\Lambda_\ell\rangle-\langle\OL{P}_\ell\grad\bdot\OL\bu_\ell\rangle$,
is constant, independent of scale $\ell$ over the intermediate range $L\gg \ell\gg\ell_\mu$.
This is simply a consequence of total energy conservation; whatever energy is input into the 
system has to either reach dissipation scales by way of the SGS flux or
get converted into internal energy through pressure dilatation.

The results of this paper lay the mathematical framework upon which we based previous 
published work \cite{Aluie11,AluieLiLi12}. This enabled us to address
basic questions pertaining to the cascade in compressible turbulence where we showed
that kinetic energy cascades conservatively despite not being in invariant of the dynamics,
and that such a cascade process is dominated by interactions between scales of similar size.

\vspace{2cm}

\noindent {\small
{\bf Acknowledgements.} 
I thank G. L. Eyink for invaluable discussions and for suggesting the idea of the proof in Proposition \ref{Prop5}
instead of a more complicated proof in an earlier version of the manuscript \cite{Aluie11arxiv}.  I also thank D. Livescu for his input on compressible and variable-density flows, S. S. Girimaji and S. K. Lele for helpful suggestions, and X. Asay-Davis for constructive comments. I wish to acknowledge the useful suggestions of two anonymous referees, and the encouragement of R. Ecke, S. Kurien, S. Li, H. Li, and B. Wingate during this project. This research was performed under the auspices of the U.S. Department of Energy at LANL under Contract No. DE-AC52-06NA25396 and supported by the LANL/LDRD program.}

\appendix
\section{Viscous effects}\label{ap:Viscouseffects}

\subsection{Constant viscous coefficient\label{ap:Viscouseffects_constant}}
In general, dynamic viscosity is a spatially varying quantity that is well-described by Sutherland's law or a simpler power law dependence on temperature, $\mu(\bx)\sim T(\bx)^{\alpha}$. Hence, it is constant in isothermal flows or approximately so in flows with a small Prandtl number where thermal conductivity is large enough to homogenize the temperature field at scales $\le \ell_\kappa$. Propositions \ref{Prop1} and \ref{Prop2} below apply to these special cases. In Propositions \ref{Prop3} and \ref{Prop4}, we generalize the proofs to flows with a spatially variable $\mu(\bx)$.

In the following proposition, we prove rigorously that viscous terms in the balance (\ref{Favremomentum}) of large-scale momentum $\OL{\rho}_\ell\wt{\bu}_\ell$ are negligible for small viscosity $\mu\to 0$. Proposition \ref{Prop1} is very similar to that given in \cite{EyinkNotes} and \cite{AluieThesis}. To avoid additional complications due to boundaries, we consider a domain $\mathbb{T}^d=[0,2\pi)^d$ that is periodic.

\begin{Prop} For a constant viscosity, $\mu(\bx)=\mu$, if velocity solutions $\bu$ of the compressible Navier-Stokes equation 
(\ref{continuity})-(\ref{internal-energy}) over a domain $\mathbb{T}^d$ have finite $2$nd-order moments:
$\int_{\mathbb{T}^d} d\bx |\bu|^2 <\infty$, 
then 
viscous terms in the large-scale momentum eq. (\ref{largemomentum}) vanish pointwise as $\mu\to 0$.
\lb{Prop1}\end{Prop}

\noindent {\it Proof of Proposition 1:}\\ 
Using integration by parts, every viscous term in eq. (\ref{largemomentum}) can be rewritten as,
\begin{eqnarray}
\mu~ \partial_j\partial_i \OL{\bu}_\ell(\bx) 
&=&\frac{\mu}{\ell^2}\int d\br (\partial_j\partial_i G)_\ell(\br) ~\bu(\bx+\br), \nonumber
\end{eqnarray}
where $(\partial_i G)_\ell(\br) = \ell^{-d}\partial G(\br/\ell)/\partial (r_i/\ell)$. This can be bounded by H\"older's inequality for $1/p + 1/q =1$,
\begin{eqnarray}
\bigg|\mu ~\partial_j\partial_i \OL{\bu}_\ell(\bx) \bigg | 
&\le& \frac{\mu}{\ell^2}\int d\br \bigg |(\partial_j\partial_iG)_\ell(\br) ~\bu(\bx+\br)    \bigg | \nonumber\\
&\le& \frac{\mu}{\ell^2} ~V^{\frac{1}{p}} ~\big\|(\partial_j\partial_iG)_\ell \big\|_p ~V^{\frac{1}{q}}\big\|\bu \big\|_q 
=\frac{\mu}{\ell^2} ~\bigg(\frac{L_{\mbox{\tiny{dom}}}}{\ell}\bigg)^{d(1-\frac{1}{p})}\big\|\bu \big\|_q 
\bigg(\int d\bs \bigg|\frac{\partial^2 G(\bs)}{\partial s_i \partial s_j} \bigg|^p \bigg)^{\frac{1}{p}}
\nonumber\end{eqnarray}
where $\|\dots\|_p = \langle|\dots|^p\rangle^{1/p}$ is the $L_p$-norm, $L_{\mbox{\tiny{dom}}}^d = V$ is the domain's volume,  and $\bs = \br/\ell$ is a dimensionless vector. Since $G(\bs)\in C^\infty$, its derivatives are uniformly bounded and we have $\big\|\partial_j\partial_iG \big\|_p = \const < \infty$ for any $p$, including $p=\infty$.
Choosing $p=q=2$, our bound implies that the viscous term $\mu \partial_j\partial_i \OL{\bu}_\ell \to 0$ at every point in space in the limit of vanishing viscosity. \hfill $\Box$
\\
\\
REMARK: We could have obtained a tighter bound in terms of $\delta u(\ell)\equiv \sup_{|\br|<\ell} |\delta\bu(\bx;\br)|$ by rewriting the viscous term as
\begin{eqnarray}
\mu~ \partial_j\partial_i \OL{\bu}_\ell(\bx) 
&=&\frac{\mu}{\ell^2}\int d\br (\partial_j\partial_i G)_\ell(\br) ~\left(\bu(\bx+\br)-\bu(\bx)\right), \nonumber
\end{eqnarray}
exploiting the fact that $\int d\br (\partial_j\partial_i G)_\ell(\br) =0$ due to the decay of $G(\bs)$ at $\pm \infty$ or due to periodic boundary conditions (see Eyink \cite{Eyink05}). Such a bound would then be of the form $O\left(\mu/\ell^2~ \delta u(\ell)\right)$ for kernels $G$ with compact support \cite{EyinkNotes}. A stronger assumption would then be required, that $\delta u(\ell)\equiv \sup_{|\br|<\ell} |\delta\bu(\bx;\br)| < \infty$ at $\bx$.

$$$$
 
The following proposition proves that viscous dissipation of large-scale kinetic energy $\frac{1}{2}\OL{\rho}_\ell |\wt{\bu}_\ell|^2$ in eq. (\ref{largeKE}) becomes negligible at any point $\bx$ in the limit of small $\mu$.
The assumptions of finite 3rd-order moments for the velocity and density fields, $\langle|\bu|^3\rangle<\infty$ and $\langle|\rho|^3\rangle<\infty$, are weak and are expected to hold in any physically realizable flow. The assumption of finite mean specific volume, $\langle1/\rho\rangle < \infty$, is used to control factors $1/\OL\rho_\ell(\bx)$ in the dissipation terms. Coarse-grained density $\OL\rho_\ell(\bx)$, for positive filter kernels $G(\br)\ge 0$, is proportional to the mass in a ball of radius $\ell$ centered around $\bx$. The assumption $\langle1/\rho\rangle < \infty$ guarantees that mass in a ball of any finite radius $\ell>0$, at any location $\bx$, will contain non-zero mass. The assumption still allows for regions with $\rho(\bx) =0$, but such regions must occupy zero volume ({\it i.e.} zero Lebesgue measure). In other words, Proposition \ref{Prop2} does not hold in vacuum pockets of non-zero volume where viscous dissipation is meaningless.

\begin{Prop}
For a constant $\mu(\bx)=\mu$, if solutions $(\rho,\bu)$ of the compressible Navier-Stokes equation (\ref{continuity})-(\ref{internal-energy}) 
over domain $\mathbb{T}^d$ have finite $3$rd-order moments:
$\int_{\mathbb{T}^d} d\bx |\rho|^3 <\infty$ and $\int_{\mathbb{T}^d} d\bx |\bu|^3 <\infty$, 
and finite mean specific volume, $\int_{\mathbb{T}^d} d\bx~ \rho^{-1} <\infty$,
then for positive kernels $G(\br)\ge 0$, viscous terms in the large-scale kinetic energy budget (\ref{largeKE}) vanish pointwise as $\mu\to 0$.
\lb{Prop2}\end{Prop}

\noindent {\it Proof of Proposition 2:}\\  
Using the exact identity $\wt\bu = \OL\bu + \OL\tau(\rho,\bu)/\OL\rho$, we have
\be \big| \partial_j\wt u_i (\bx) \big|
\le \big| \partial_j\OL u_i \big| + \big| \frac{1}{\OL\rho}\partial_j\OL\tau(\rho,u_i)\big|
+ \big|  \frac{1}{\OL\rho^2}\OL\tau(\rho,u_i)\partial_j\OL\rho \big|.
\lb{prop2_1}\ee

The first term $\big| \partial_j\OL u_i \big|$ is bounded by 
\begin{eqnarray}
&&\ell^{-1}\|\bu \|_3 ~\const\| (\grad G)_\ell \|_2  \nonumber\\
&=&\ell^{-1} \|\bu \|_3 ~A(L{\mbox{\tiny{dom}}}/\ell),
\lb{prop2_bound1}\end{eqnarray}
through an argument identical to that in Proposition \ref{Prop1} and using the fact that 
$\|\bu \|_2\le \const\|\bu \|_3$ over a bounded domain. Here, $A(L{\mbox{\tiny{dom}}}/\ell) = \const(L{\mbox{\tiny{dom}}}/\ell)^{d/2}\| \grad G \|_2$ is dimensionless. 

The second term in (\ref{prop2_1}) can be rewritten as
$$\frac{1}{\OL\rho}\partial_j\OL\tau(\rho,u_i)
=-\frac{1}{\OL\rho\ell}\bigg[\int d\br (\partial_j G)_\ell(\br) \rho(\bx+\br) u_i(\bx+\br)$$
$$-\int d\br (\partial_j G)_\ell(\br) \rho(\bx+\br)\int d\br' G_\ell(\br') u_i(\bx+\br')
-\int d\br G_\ell(\br) \rho(\bx+\br)\int d\br' (\partial_j G)_\ell(\br') u_i(\bx+\br')\bigg],
$$
using integration by parts. Employing the 3-3-3 H\"older's inequality, this expression is bounded by
\begin{eqnarray}
&&\const \ell^{-1} \frac{1}{\OL{\rho}}  \big\|\rho  \big\|_3\big\|\bu  \big\|_3 (1+2\big\| 1 \big\|^2_3\big\| G_\ell \big\|_3) \big\|(\nabla G)_\ell \big\|_3 \nonumber\\
&=&\ell^{-1} \frac{1}{\OL{\rho}}  \big\|\rho  \big\|_3\big\|\bu  \big\|_3 ~B(L{\mbox{\tiny{dom}}}/\ell)
\lb{prop2_bound2}\end{eqnarray}
where $B(L{\mbox{\tiny{dom}}}/\ell) = \left(\const \left(\frac{L{\mbox{\tiny{dom}}}}{\ell}\right)^{\frac{2d}{3}}+\const \left(\frac{L{\mbox{\tiny{dom}}}}{\ell}\right)^{\frac{4d}{3}}\right)$ is dimensionless.

The third term in (\ref{prop2_1}) can be rewritten as
$$ \frac{1}{\OL\rho^2}\OL\tau(\rho,u_i)\partial_j\OL\rho
=-\OL\rho^{-2}\ell^{-1} \int d\br (\partial_j G)_\ell(\br) \rho(\bx+\br)$$
$$\times\bigg[\int d\br G_\ell(\br) \rho(\bx+\br) u_i(\bx+\br)
-\int d\br G_\ell(\br) \rho(\bx+\br)\int d\br' G_\ell(\br') u_i(\bx+\br') \bigg].$$
Using H\"older's inequality, this is bounded by
\begin{eqnarray}
&&\const \ell^{-1} \frac{1}{\OL{\rho}^2} \big\|\rho  \big\|^2_3\big\|\bu  \big\|_3
\left(1+\big\|1 \big\|_3\big\| G_\ell \big\|_3\right) \big\|(\nabla G)_\ell \big\|_3 \big\|1 \big\|_3 \big\|G_\ell \big\|_3 \nonumber\\
&=& \ell^{-1} \frac{1}{\OL{\rho}^2} \big\|\rho  \big\|^2_3\big\|\bu  \big\|_3 ~C(L{\mbox{\tiny{dom}}}/\ell)
\lb{prop2_bound3}\end{eqnarray}
where $C(L{\mbox{\tiny{dom}}}/\ell) = \left(\const \left(\frac{L{\mbox{\tiny{dom}}}}{\ell}\right)^{\frac{4d}{3}} + \const \left(\frac{L{\mbox{\tiny{dom}}}}{\ell}\right)^{2d}\right)$ is dimensionless.

Finally, each of the viscous terms is bounded by
$$\mu \big| \grad\wt \bu \grad\OL{\bu} \big|
\le \frac{\mu}{\ell^{2}}\big\|\bu  \big\|^2_3
\bigg[ A(L{\mbox{\tiny{dom}}}/\ell) + B(L{\mbox{\tiny{dom}}}/\ell) \frac{\big\|\rho  \big\|_3}{\OL\rho}+ C(L{\mbox{\tiny{dom}}}/\ell) \frac{\big\|\rho  \big\|^2_3}{\OL\rho^2}\bigg]A(L{\mbox{\tiny{dom}}}/\ell).
$$
Factors $1/\OL\rho_\ell (\bx)$ in the above expression are finite because $1/\rho$ is a convex function of density over $\rho \in [0,\infty)$. When $G(\br)\ge 0$, coarse-graining is an averaging operation and we can use Jensen's inequality to obtain 
\begin{eqnarray}
1/\OL\rho_\ell (\bx) \le (\OL{1/\rho})_\ell (\bx) \le \| G_\ell \|_p \| \rho^{-1} \|_q = \| \rho^{-1} \|_q \left(\frac{L{\mbox{\tiny{dom}}}}{\ell}\right)^{d(1-\frac{1}{p})} \| G \|_p, 
\nonumber\end{eqnarray}
where we used H\"older's inequality to obtaining the second inequality with $1/p + 1/q =1$. For $p=\infty$ and $q=1$, we have that $1/\OL\rho_\ell (\bx) < \infty$ for any fixed $\ell>0$ under the assumption $\langle1/\rho\rangle<\infty$.

Hence, for any fixed $\ell>0$, a viscous term $\mu \grad\wt \bu \grad\OL{\bu}  \to 0$ at every point in space in 
the limit of vanishing viscosity.
 \hfill $\Box$

$$$$

\subsection{Spatially varying viscous coefficient\label{ap:Viscouseffects_variable}}
We now extend the previous two propositions to the case when $\mu(\bx)$ varies in space.
In the proofs below, we require that spatial gradients of viscosity are bounded (in a root-mean-square sense) and vanish when $\mu_{\mathrm{rms}}\to 0$. The assumption can be checked directly from a series of direct numerical simulations at increasingly higher resolution. Empirical support (or refutation) for the assumption can also be obtained by measuring the spectrum of $\mu(\bx)$, namely $E^\mu(k)=\sum_{k-1<|\bk|\le k} |\widehat{\mu}(\bk)|^2$ or, alternatively, by measuring the temperature spectrum based on a relation $\mu = \mu(T)$. 

A spectrum $E^\mu(k)$ that decays faster than $k^{-1/3}$ would support to our assumption on $\grad\mu$ based on the following physical reasoning.  $\mu(T(\bx))$ varies on scales $\ge \ell_\kappa$, the temperature dissipation scale. If such variations scale as $\delta \mu(\ell) \sim \mu (\ell/L)^{h_\mu}$ for $\ell\gg \ell_\kappa$, corresponding to a spectrum $E^\mu(k)\sim k^{-2h_\mu-1}$, then gradients scale as 
$$\grad\mu \sim \delta \mu(\ell)/\ell \sim \mu (\ell/L)^{h_\mu-1}.$$
If $h_\mu \ge 1$, the field $\mu(\bx)$ is smooth enough to guarantee that gradients have little contribution from small scales and $\grad\mu$ will vanish as $\mu\to 0$. This is also true if $E^\mu(k)$ decays faster than any power law (e.g. exponentially). On the other hand, if $h_\mu < 1$, most of the contribution to $\grad \mu$ comes from the smallest scales in the temperature field and we have 
$$\grad\mu \sim \mu (\ell_\kappa/L)^{h_\mu-1}.$$
Suppose the temperature dissipation scale, $\ell_\kappa$, can be determined from the Obukhov-Corrsin-Batchelor phenomenology of passive scalars \cite{,Obukhov49,Corrsin51,Batchelor59}:
\begin{eqnarray}
\ell_\kappa&=&\left(\frac{\kappa^3}{\langle\rho\rangle^2\epsilon_\mathrm{net} ~c_p^3}\right)^{\frac{1}{4}},  \hspace{3.1cm}\mbox{for~} Pr\le 1 \lb{OCB_smallPr}\\
\ell_\kappa&=&\left(\frac{\kappa^2 ~\mu}{\langle\rho\rangle^2\epsilon_\mathrm{net} ~c_p^2}\right)^{\frac{1}{4}} = \ell_\mu~Pr^{-\frac{1}{2}},  \hspace{1.2cm} \mbox{for~} Pr > 1, \lb{OCB_largePr}
\end{eqnarray}
where $\epsilon_\mathrm{net}= \langle\rho\bu\bdot{\bf F} + P\grad\bdot\bu\rangle$ is the net kinetic energy reaching the dissipation scales (see eq. (10) in Aluie \cite{Aluie11} or eq. (\ref{steadyKE_II}) above).
The Prandtl number is $Pr = c_p \mu/\kappa$, where $c_p$ is specific heat at constant pressure. The second equality in eq. (\ref{OCB_largePr}) assumes that the kinetic dissipation scale is determined from a Kolmogorov-type estimate 
$\ell_\mu = (\mu^3/\langle\rho\rangle^2\epsilon_\mathrm{net} )^{1/4}$.

It follows from relations (\ref{OCB_smallPr}),(\ref{OCB_largePr}) that viscosity gradients vanish when $\mu\to 0$ for a fixed $Pr$ ({\it i.e.} in the limit of small $\mu$ and small $\kappa$ while keeping their ratio constant), 
\begin{eqnarray}
\grad\mu &\sim& \mu (\ell_\kappa/L)^{h_\mu-1} \sim \mu^{\frac{3h_\mu+1}{4}} Pr^{\frac{3}{4}(1-h_\mu)},  \hspace{1cm}\mbox{for~} Pr\le 1,\nonumber\\[0.3cm]
\grad\mu &\sim& \mu (\ell_\kappa/L)^{h_\mu-1} \sim \mu^{\frac{3h_\mu+1}{4}} Pr^{\frac{1-h_\mu}{2}},  \hspace{1.5cm}\mbox{for~} Pr> 1,\nonumber
\end{eqnarray}
under the very weak condition that $h_\mu>-1/3$. If, on the other hand, we consider the limit $\mu\to 0$ for a fixed $\kappa$ ({\it i.e.} in the limit of small $\mu$ and small $Pr$), then $\grad\mu \sim \mu$ vanishes for any value of the scaling exponent $h_\mu$, as one would expect physically for flows in the $Pr\to 0$ limit where thermal conductivity becomes large enough to homogenize the temperature field.

Measuring $E^\mu(k)$ from direct numerical simulations is simple although we are not aware of any such result reported in the literature. We note that a decay rate of $E^\mu(k)< \const k^{-1/3}$ is a weak condition on any reasonable field. Alternatively, since viscosity is related to temperature through Sutherland's law or a simpler power law, $\mu(\bx) = \mu_0 (T(\bx)/T_0)^{\alpha}$, one can infer the spectrum $E^\mu(k)$ from that of temperature, $E^T(k)$. This follows from $\mu(T)$ being a Lipschitz function of temperature and $T(\bx)\ge T_\mathrm{min}>0$:
$$ \delta \mu(\ell) =  \mu(x+\ell)- \mu(x) 
= \frac{\mu_0}{T_0^\alpha}\,\alpha\, T_\mathrm{cst}^{\alpha-1} [T(x+\ell) - T(x)] 
= \const \mu_0  \left(\frac{\ell}{L}\right)^{h_T},
$$
where we used the mean value theorem in the second equality with $T_\mathrm{cst} \in (T(x),T(x+\ell))$, and 
$\delta T(\ell)=\const T_\mathrm{rms} (\ell/L)^{h_T}$ in the last equality. Hence, fluctuations in $\mu$ have the same scaling exponent as those of the temperature field, $h_\mu = h_T$.  We note that the physical arguments presented thus far in this subsection \ref{ap:Viscouseffects_variable} are not rigorous but serve to justify the assumption on $\grad\mu$ used in the rigorous proofs of Propositions \ref{Prop3} and \ref{Prop4}.

$$$$

\begin{Prop} If velocity solutions $\bu$ of the compressible Navier-Stokes equation 
(\ref{continuity})-(\ref{internal-energy}) over a domain $\mathbb{T}^d$ have finite $2$nd-order moments:
$\int_{\mathbb{T}^d} d\bx |\bu|^2 <\infty$, and viscous gradients are bounded as follows: 
$$\langle|\grad\mu(\bx)|^2\rangle^{\frac{1}{2}} \le A~ \frac{\mu_{\mathrm{rms}}}{L} ~\left(\frac{\mu_{\mathrm{rms}}}{M}\right)^{\frac{3}{4}(h_\mu-1)},$$ 
where $\mu_{\mathrm{rms}} = \langle\mu^2\rangle^{1/2}$, $\beta = 3/4(h_\mu-1)>-1$  and constants $A$, $L$, and $M$, then viscous terms in the large-scale momentum eq. (\ref{largemomentum}) vanish pointwise as $\mu_{\mathrm{rms}}\to 0$.
\lb{Prop3}\end{Prop}

\noindent {\it Proof of Proposition 3:}\\  
By integration by parts, any of the viscous terms in eq. (\ref{largemomentum}) can be rewritten as,
\begin{eqnarray}
\partial_j (\OL{\mu ~\partial_i \bu})_\ell(\bx)  
&=&\frac{1}{\ell^2}\int d\br (\partial_i \partial_j G)_\ell(\br) ~\mu(\bx+\br)~ \bu(\bx+\br) \nonumber\\
&+&\frac{1}{\ell}\int d\br (\partial_j G)_\ell(\br) ~\bu(\bx+\br)~\partial_i\mu(\bx+\br).\nonumber
\end{eqnarray}
This can be bounded by
\begin{eqnarray}
\bigg|\partial_j (\OL{\mu ~\partial_i \bu})_\ell(\bx) \bigg| 
&\le& \frac{1}{\ell^2} \big\|(\partial_i \partial_jG)_\ell \big\|_\infty \big\|\mu \big\|_2\big\|\bu \big\|_2 + \frac{1}{\ell} \big\|(\partial_j G)_\ell \big\|_\infty \big\|  \bu \big\|_2 \big\|\grad \mu \big\|_2 \nonumber\\
&\le& \frac{\mu_{\mathrm{rms}}}{\ell^2} \big\|\bu \big\|_2 \left(\frac{L_{\mathrm{dom}}}{\ell}\right)^d  \big\|\partial_i \partial_jG \big\|_\infty  + \frac{\mu_{\mathrm{rms}}^{1+\beta}}{\ell ~L ~M^{\beta}} ~A ~   \big\|  \bu   \big\|_2  \left(\frac{L_{\mathrm{dom}}}{\ell}\right)^d \big\|\partial_j G \big\|_\infty ,
\nonumber\end{eqnarray}
which, for any fixed $\ell > 0$, vanishes as $\mu_{\mathrm{rms}}\to 0$.
 \hfill $\Box$
\\
\\
REMARK: In our choice of $\mu_{\mathrm{rms}}$ when deriving the above bound, we imagined a field $\mu(\bx)$ that is statistically homogeneous. If $\mu(\bx)$ in a region of interest is significantly different (in a statistical sense) from the rest of the domain, other more optimal bounds can be derived using other moments of $\mu(\bx)$ such as $\|\mu\|_{L^2_{loc}}$ or $\max_{|\bx+\br|<\ell}\mu(\bx+\br)$.

$$$$

\begin{Prop}
If solutions $(\rho,\bu)$ of the compressible Navier-Stokes equation (\ref{continuity})-(\ref{internal-energy}) 
over domain $\mathbb{T}^d$ have finite $3$rd-order moments:
$\int_{\mathbb{T}^d} d\bx |\rho|^3 <\infty$ and $\int_{\mathbb{T}^d} d\bx |\bu|^3 <\infty$, 
a finite mean specific volume, $\int_{\mathbb{T}^d} d\bx~ \rho^{-1} <\infty$, and viscous gradients are bounded as follows:  
$$\langle|\grad\mu(\bx)|^2\rangle^{\frac{1}{2}} \le A~ \frac{\mu_{\mathrm{rms}}}{L} ~\left(\frac{\mu_{\mathrm{rms}}}{M}\right)^{\frac{3}{4}(h_\mu-1)},$$ 
where $\mu_{\mathrm{rms}} = \langle\mu^2\rangle^{1/2}$, $\beta = 3/4(h_\mu-1)>-1$  and constants $A$, $L$, and $M$, then for positive kernels $G(\br)\ge 0$, viscous terms in the large-scale kinetic energy budget (\ref{largeKE}) vanish pointwise as $\mu_{\mathrm{rms}}\to 0$.
\lb{Prop4}\end{Prop}

\noindent {\it Proof of Proposition 4:}\\  
As in Proposition \ref{Prop2}, eqs. (\ref{prop2_1})-(\ref{prop2_bound3}), we have the following bound on $\grad\wt\bu$:
$$\big| \grad\wt\bu \big|
\le \frac{1}{\ell}\big\|\bu  \big\|_3
\bigg[ A(L{\mbox{\tiny{dom}}}/\ell) + B(L{\mbox{\tiny{dom}}}/\ell) \frac{\big\|\rho  \big\|_3}{\OL\rho}+ C(L{\mbox{\tiny{dom}}}/\ell) \frac{\big\|\rho  \big\|^2_3}{\OL\rho^2}\bigg].
$$

In a derivation similar to that in Proposition \ref{Prop3}, we obtain the following bound
\begin{eqnarray}
\bigg|\OL{\mu ~ \grad \bu}(\bx) \bigg| 
&\le& \frac{1}{\ell} \big\|(\grad G)_\ell \big\|_\infty \big\|\mu \big\|_2\big\|\bu \big\|_2 +  \big\|G_\ell \big\|_\infty \big\|  \bu \big\|_2 \big\|\grad \mu \big\|_2 \nonumber\\
&\le& \const \frac{\mu_{\mathrm{rms}}}{\ell} \big\|\bu \big\|_3 \left(\frac{L_{\mathrm{dom}}}{\ell}\right)^d  \big\|\grad G \big\|_\infty  + \const \frac{\mu_{\mathrm{rms}}^{1+\beta}}{L ~M^{\beta}}    \big\|  \bu   \big\|_3  \left(\frac{L_{\mathrm{dom}}}{\ell}\right)^d \big\| G \big\|_\infty ,
\nonumber\end{eqnarray}
Hence, we have 
\begin{eqnarray}
\bigg|\grad\wt\bu ~\OL{\mu ~ \grad \bu}(\bx) \bigg| 
\le \frac{\mu_{\mathrm{rms}}}{\ell^2}\big\|\bu  \big\|^2_3
&&\bigg[ A(L{\mbox{\tiny{dom}}}/\ell) + B(L{\mbox{\tiny{dom}}}/\ell) \frac{\big\|\rho  \big\|_3}{\OL\rho}+ C(L{\mbox{\tiny{dom}}}/\ell) \frac{\big\|\rho  \big\|^2_3}{\OL\rho^2}\bigg]\nonumber\\
&\times& \bigg[ \const + \const \frac{\ell}{L}\frac{\mu_{\mathrm{rms}}^{\beta}}{M^{\beta}}   \bigg] \left(\frac{L_{\mathrm{dom}}}{\ell}\right)^d
\nonumber\end{eqnarray}
As discussed in Proposition \ref{Prop2}, factors $1/\OL\rho_\ell (\bx)$ are bounded by 
\begin{eqnarray}
1/\OL\rho_\ell (\bx) \le (\OL{1/\rho})_\ell (\bx) \le \| G_\ell \|_p \| \rho^{-1} \|_q = \| \rho^{-1} \|_q \left(\frac{L{\mbox{\tiny{dom}}}}{\ell}\right)^{d(1-\frac{1}{p})} \| G \|_p. 
\nonumber\end{eqnarray}
It follows that for any fixed $\ell>0$, viscous dissipation in eq. (\ref{largeKE}), $D_\ell(\bx) \to 0$, at every $\bx$ in the limit  $\mu_{\mathrm{rms}}\to 0$.
 \hfill $\Box$

$$$$

\section{Kinetic energy injection}\label{ap:Kineticenergyinjection}

The following proposition proves that mean kinetic energy injection is localized to the largest scales $\gtrsim L\sim K_0^{-1}$ when stirring with an external acceleration field. To this end, we need to show that 
\be\langle\epsilon^{inj}_\ell\rangle = \langle \OL\rho\, \wt{u}_i \wt{F}_i \rangle
\longrightarrow \big\langle \left({\rho u_i}\right)^{<K_0} {F}_i^{<K_0} \big\rangle,\ee 
as $K_0\ell \to 0$.
In our proof, we will assume that the external acceleration field has only small wavenumber modes $<K_0$ as is traditionally done in numerical simulations (e.g. \cite{KidaOrszag90a,Schmidtetal09,Federrathetal10,Kritsuketal07}) such that ${\bf F}(\bx)={\bf F}^{<K_0}(\bx)$ for some $K_0>0$ (the case of $K_0=0$ was discussed in section \ref{sec:RTflow}). We will also assume that density fluctuations
decay fast enough (or at least grow slowly) at small scales such that for some constant $A_2$, $\|\delta \rho(\br)\|_2 \le \rho_{rms} A_2 (rK_0)^{\sigma_2^\rho}$ with $\sigma_2^\rho > -1$ as $r\to 0$. This condition corresponds to 
a density spectrum $E^{\rho}(k)\equiv\sum_{k-0.5<|\bk|< k+0.5} |\hat\rho(\bk)|^2 \le \const k^{-\alpha}$ 
with $\alpha = 2\sigma_2^\rho+1 > -1$, which is readily satisfied for any density field with finite $\rho_{rms}$.
Such an assumption is only an upper bound and does not even require that the density spectrum be continuous .
We also assume that $\langle1/\rho^4\rangle < \infty$, which is slightly stronger that an assumption of finite mean specific volume used in Proposition \ref{Prop2}. Regarding the filtering kernel, we assume
that it decays fast enough with $r\to\infty$ such that $\int d\br \,\, |G(\br)| |\br|^{\beta} =\const< \infty$ for 
$\beta = 1$, $\sigma_2^\rho$, and $1+\sigma_2^\rho$, which hold for any reasonable kernel with finite spread.

\begin{Prop} 
If solutions $(\rho,\bu)$ of the compressible Navier-Stokes equation (\ref{continuity})-(\ref{internal-energy}), 
stirred with an acceleration ${\bf F}(\bx) = {\bf F}^{<K_0}(\bx)$ in a periodic domain ${\mathbb{T}^d}=[0,2\pi)^d$, 
satisfy
$ \|\rho\bu\|_4<\infty$, $ \|1/\rho\|_4<\infty$, and  $\|\delta\rho(\br)\|_2\le\rho_{rms}A_2 (rK_0)^{\sigma^\rho_2}$, then
\begin{eqnarray}
&&\hspace{-1cm}\big|\langle\epsilon^{inj}_\ell - \left({\rho u_i}\right)^{<K_0}   {F}_i^{<K_0} \rangle\big|    \nonumber\\
&\le& \const   \rho_{rms}\|\rho^{-1}\|_4\|\rho\bu\|_4\|{\bf F}\|_1  \left(\ell K_0\right)^{1+\sigma^\rho_2}
+    \const  \|\rho\bu\|_1\|{\bf F}\|_1  \left(\ell K_0\right)^{2}
\hspace{2cm}\lb{injectionconv}\end{eqnarray}
when $G(\br)\ge 0$ and $\int d\br \,\, |G(\br)| |\br|^{\beta} =\const< \infty$ for $\beta =1$, $\sigma_2^\rho$, and $1+\sigma_2^\rho$.
The upper bound (\ref{injectionconv}) vanishes in the limit $\ell K_0\to 0$ for any $\sigma^\rho_2 > -1$.
\lb{Prop5}\end{Prop}

\noindent {\it Proof of Proposition 5:}\\ 
Rewriting the kinetic energy injection as
$$\epsilon^{inj}_\ell \equiv \OL\rho\,\wt{u}_i\wt{F}_i = \frac{\OL{\rho u_i}}{\OL\rho}\OL{\rho F_i} 
=\frac{\OL{\rho u_i}}{\OL\rho}\left[\OL\rho\,\, \OL{F}_i + \OL\tau(\rho,F_i)\right],
$$
we have 
\begin{eqnarray}
&&\langle\epsilon^{inj}_\ell -(\rho u_i)^{<K_0} F_i^{<K_0} \rangle \nonumber\\
&=& \left\langle\OL{\rho u_i}\OL{F}_i - (\OL{\rho u_i})^{<K_0} \left(\OL{F}_i\right)^{<K_0} \right\rangle
+ \left\langle\frac{\OL{\rho u_i}}{\OL\rho} \OL\tau(\rho,F_i) \right\rangle
- \left\langle \OL\tau\left((\rho u_i)^{<K_0},~ F_i^{<K_0} \right) \right\rangle,
\nonumber\end{eqnarray}
where we used $\int d\bx~ f(\bx) = \int d\bx~ \OL{f}(\bx)$ in the last term. 
Due to commutativity and associativity of convolutions, $\OL{\left({\bf F}^{<K_0}\right)} = \left(\OL{\bf F}\right)^{<K_0}$,
the orthogonality of Fourier modes, $\langle\OL{\rho u_i} \left(\OL{F}_i\right)^{<K_0}\rangle = \langle\left(\OL{\rho u_i}\right)^{<K_0}\OL{F}_i^{<K_0}\rangle$, yields
\begin{eqnarray}
\big|\langle\epsilon^{inj}_\ell - \left({\rho u_i}\right)^{<K_0} {F}_i^{<K_0} \rangle\big|
&=& \big|\langle\frac{\OL{\rho u_i}}{\OL\rho} \OL\tau(\rho,F_i) \rangle  - \left\langle \OL\tau\left((\rho u_i)^{<K_0},~ F_i^{<K_0} \right) \right\rangle\big| \nonumber\\
&\le& \bigg\|  \frac{\OL{\rho u_i}}{\OL\rho} \OL\tau(\rho,F_i) \bigg\|_1 +
\big\|  \OL\tau\left((\rho u_i)^{<K_0},~ F_i^{<K_0} \right) \big\|_1  \nonumber\\
&\le& \bigg\|\frac{1}{\OL\rho} \bigg\|_4  \big\|\rho \bu  \big\|_4  \big\|\OL\tau(\rho,{\bf F}) \big\|_2  \int d\br\,\,|G_\ell(\br)|\nonumber\\
&&\hspace{-.3cm}+ \big\|  \OL\tau\left((\rho u_i)^{<K_0},~ F_i^{<K_0} \right) \big\|_1 \hspace{1cm}
\lb{app:injectionbound1}\end{eqnarray}
We have $\int d\br\,\,|G_\ell(\br)| = 1$ when $G(\br)\ge 0$. Furthermore, observe that $1/\rho$ is a convex function of density over $\rho \in [0,\infty)$ such that when $G(\br)\ge 0$, coarse-graining is an averaging operation and we can use Jensen's inequality to obtain $1/\OL\rho_\ell \le (\OL{1/\rho})_\ell $. It follows from (\ref{app:injectionbound1}) that 
\begin{eqnarray}
 \big|\langle\epsilon^{inj}_\ell &-& \left({\rho u_i}\right)^{<K_0} {F}_i^{<K_0} \rangle\big|\nonumber\\
&\le& \bigg\|\frac{1}{\rho} \bigg\|_4  \big\|\rho \bu  \big\|_4  \big\|\OL\tau(\rho,{\bf F}) \big\|_2
+ \big\|  \OL\tau\left((\rho u_i)^{<K_0},~ F_i^{<K_0} \right) \big\|_1.
\lb{app:injectionbound2}\end{eqnarray}
It is straightforward to verify the following identity due to \cite{Constantinetal94,Eyink95a},
\begin{eqnarray}
\OL\tau(f,g)(\bx) &=& \int d\br \,\,G_\ell(\br) \delta f(\br;\bx) \delta g(\br;\bx)\nonumber\\
&&- \int d\br_1 G_\ell(\br_1) \delta f(\br_1;\bx) \int d\br_2 G_\ell(\br_2) \delta g(\br_2;\bx), 
\hspace{1cm} \lb{ap:identityforcing}\end{eqnarray}
where $ \delta f(\br;\bx) = f(\bx+\br) - f(\bx)$. Identity (\ref{ap:identityforcing}) allows us to 
derive the following upper bound:
\begin{eqnarray}
\big\| \OL\tau(\rho,{\bf F})(\bx)\big\|_2 &\le& \int d\br \,\,|G_\ell(\br)| \|\delta\rho(\br;\bx)\|_2 ~\|\delta {\bf F}(\br;\bx)\|_\infty\nonumber\\
&&+ \int d\br_1 |G_\ell(\br_1)| \| \delta\rho(\br_1;\bx)\|_2 \int d\br_2 |G_\ell(\br_2)| \|\delta {\bf F}(\br_2;\bx) \|_\infty\nonumber\\
&\le& \const \int d\br \,\,|G_\ell(\br)| ~(|\br| K_0)^{\sigma_2^{\rho}+1}  \nonumber\\
&&+ \const\int d\br_1 |G_\ell(\br_1)| ~(|\br_1| K_0)^{\sigma_2^{\rho}}  \int d\br_2 |G_\ell(\br_2)|~(|\br_2| K_0) \nonumber\\
&=& (\ell K_0)^{\sigma_2^{\rho}+1}
\bigg\{\const \int d\br \,\,|G(\br)| ~|\br|^{\sigma_2^{\rho}+1} \nonumber\\
&&\hspace{2cm}+ \const\int d\br_1 |G(\br_1)| ~|\br_1|^{\sigma_2^{\rho}}  \int d\br_2 |G(\br_2)|~|\br_2| \bigg\}\nonumber\\
&=& O\left(\left(\ell K_0\right)^{\sigma_2^{\rho}+1}\right),
\hspace{9cm}\lb{app:injectionbound3}\end{eqnarray}
where we used ${\bf F}(\bx) = {\bf F}^{<K_0}(\bx)$ to infer the uniform Lipschitz condition,
$|\delta {\bf F} (\br;\bx)| \le \const \|{\bf F}\|_1K_0|\br|$. 
Using identity (\ref{ap:identityforcing}), we also have
\begin{eqnarray}
\big\|  \OL\tau\left((\rho u_i)^{<K_0},~ F_i^{<K_0} \right) \big\|_1 
= O\left(\left(\ell K_0\right)^{2}\right),
\lb{app:injectionbound4}\end{eqnarray} due to steps similar to those leading to (\ref{app:injectionbound3}).
Result (\ref{injectionconv}) follows from bounds (\ref{app:injectionbound2}), (\ref{app:injectionbound3}), and (\ref{app:injectionbound4}).
 \hfill $\Box$
 
 $$$$

The following proposition proves that mean kinetic energy injection is localized to the largest scales $\gtrsim L\sim K_0^{-1}$ when stirring with an external forcing field. To this end, we need to show that 
\be\langle\epsilon^{inj}_\ell\rangle = \langle \, \wt{u}_i \OL{{\mathcal F}}_i \rangle
\longrightarrow \big\langle u_i^{<K_0} {{\mathcal F}}_i^{<K_0} \big\rangle,\ee 
as $K_0\ell \to 0$. The bound  below is weaker than that obtained in Proposition \ref{Prop5} because $\sigma^{u}_4 < 1$ in a turbulent flow. For example, $\sigma^{u}_4 = 1/3$ in Kolmogorov's 1941 theory. This suggest that the rate of convergence is slower than that in Proposition \ref{Prop5}, and that kinetic energy injection is less localized when stirring with ${\boldsymbol {\mathcal F}}^{<K_0}$ compared to when stirring with $\rho{\bf F}^{<K_0}$. This is certainly true when $K_0 = 0$, in which case stirring with $\rho{\bf F}_\mathrm{const}$ injects energy only at the largest scale, $L_\mathrm{dom}$, as discussed in section \ref{sec:RTflow} above. On the other hand stirring with ${\boldsymbol {\mathcal F}}_\mathrm{const}$ yields an energy injection
$\langle\epsilon^{inj}_\ell\rangle = {\boldsymbol {\mathcal F}}_\mathrm{const}\bdot\langle\wt\bu_\ell\rangle
= {\boldsymbol {\mathcal F}}_\mathrm{const}\bdot\langle\OL\tau_\ell(\rho,\bu)/\OL\rho_\ell\rangle$, which in general is non-zero at scales $<L_\mathrm{dom}$.
 
 \begin{Prop} 
If solutions $(\rho,\bu)$ of the compressible Navier-Stokes equation (\ref{continuity})-(\ref{internal-energy}), 
stirred with a force ${\boldsymbol {\mathcal F}}(\bx)={\boldsymbol {\mathcal F}}^{<K_0}(\bx)$ in a periodic domain ${\mathbb{T}^d}=[0,2\pi)^d$, satisfy
$ \|1/\rho\|_4<\infty$,  $\|\delta\rho(\br)\|_4\le\rho_{rms}A_4 (rK_0)^{\sigma^\rho_4}$, and
$\|\delta\bu(\br)\|_4\le u_{rms}B_4 (rK_0)^{\sigma^u_4}$, then
\begin{eqnarray}
&&\hspace{-1cm}\big|\langle\epsilon^{inj}_\ell - u_i^{<K_0}   {\mathcal F}_i^{<K_0} \rangle\big|    \nonumber\\
&\le& \const   \rho_{rms}u_{rms}\|\rho^{-1}\|_4\|{\boldsymbol {\mathcal F}}\|_4  \left(\ell K_0\right)^{\sigma^\rho_4+\sigma^u_4}
+    \const  \|\bu\|_1\|{\boldsymbol {\mathcal F}}\|_1  \left(\ell K_0\right)^{2}
\hspace{2cm}\lb{injectionconv_force}\end{eqnarray}
when $G(\br)\ge 0$ and $\int d\br \,\, |G(\br)| |\br|^{\beta} =\const< \infty$ for $\beta =\sigma_4^\rho$,  $\sigma_4^u$, and $\sigma_4^\rho+\sigma_4^u$.
The upper bound (\ref{injectionconv_force}) vanishes in the limit $\ell K_0\to 0$ for any $\sigma^\rho_4+\sigma^u_4 > 0$.
\lb{Prop6}\end{Prop}

\noindent {\it Proof of Proposition 6:}\\ 
Rewriting the kinetic energy injection as
$$\epsilon^{inj}_\ell \equiv  \wt{u}_i \OL{{\mathcal F}}_i = \frac{\OL{\rho u_i}}{\OL\rho}\OL{{\mathcal F}}_i
=\left[\frac{\OL\rho~\OL{u}_i}{\OL\rho} + \frac{\OL\tau(\rho,u_i)}{\OL\rho} \right]\OL{{\mathcal F}}_i,
$$
we have 
\begin{eqnarray}
&&\langle\epsilon^{inj}_\ell -u_i^{<K_0} {\mathcal F}_i^{<K_0} \rangle \nonumber\\
&=& \left\langle\OL{u_i}~\OL{{\mathcal F}}_i - (\OL{u_i})^{<K_0} \left(\OL{{\mathcal F}}_i\right)^{<K_0} \right\rangle
+ \left\langle\frac{\OL\tau(\rho,u_i)}{\OL\rho} \OL{{\mathcal F}}_i \right\rangle
- \left\langle \OL\tau\left(u_i^{<K_0},~ {\mathcal F}_i^{<K_0} \right) \right\rangle,
\nonumber\end{eqnarray}
where we used $\int d\bx~ f(\bx) = \int d\bx~ \OL{f}(\bx)$ in the last term. 
Orthogonality of Fourier modes, $\langle\OL{u_i} \left(\OL{{\mathcal F}}_i\right)^{<K_0}\rangle = \langle\left(\OL{u_i}\right)^{<K_0}\left(\OL{{\mathcal F}}_i\right)^{<K_0}\rangle$, yields
\begin{eqnarray}
\big|\langle\epsilon^{inj}_\ell  -u_i^{<K_0} {\mathcal F}_i^{<K_0}  \rangle\big|
&=& \big|\left\langle\frac{\OL\tau(\rho,u_i)}{\OL\rho} \OL{{\mathcal F}}_i \right\rangle
- \left\langle \OL\tau\left(u_i^{<K_0},~ {\mathcal F}_i^{<K_0} \right) \right\rangle\big| \nonumber\\
&\le& \const \bigg\|\frac{1}{\rho} \bigg\|_4  \big\|{\boldsymbol {\mathcal F}}  \big\|_4  \big\|\OL\tau(\rho,u_i) \big\|_2  \nonumber\\
&&\hspace{-.3cm}+ \big\|  \OL\tau\left(u_i^{<K_0},~ {\mathcal F}_i^{<K_0} \right) \big\|_1 \hspace{1cm}
\lb{app:injectionbound1}\end{eqnarray}

Similar to bounds (\ref{app:injectionbound3}) and (\ref{app:injectionbound4}) in Proposition \ref{Prop5}, we have 
\begin{eqnarray}
\big\|\OL\tau(\rho,u_i) \big\|_2  &=& O\left(\left(\ell K_0\right)^{\sigma_4^{\rho}+\sigma_4^{u}}\right)\\
 \big\|  \OL\tau\left(u_i^{<K_0},~ {\mathcal F}_i^{<K_0} \right) \big\|_1&=& O\left(\left(\ell K_0\right)^{2}\right),
\end{eqnarray}
Result (\ref{injectionconv_force}) follows.
\hfill $\Box$

$$$$

\section{Small-scale kinetic energy budget}\label{ap:SmallKE}
For completeness, we derive the kinetic energy budget at scales $<\ell$ which complements
large-scale budget (\ref{largeKE}). A small-scale kinetic energy may be defined as 
\be
\wt{k}_\ell \equiv \OL\rho_\ell \frac{\wt\tau_\ell(u_i,u_i)}{2}.
\nonumber\ee
Integrating $\wt{k}_\ell$ in space gives $\int d\bx\, \rho |\bu|^2/2-\int d\bx\, \OL\rho |\wt\bu|^2/2$, which is
the total kinetic energy less the energy at large scales. Similar to the large-scale budget (\ref{largeKE}), it is straightforward to derive a budget for $\wt{k}_\ell$, which reads
\be
\partial_t \frac{\OL\rho_\ell\wt\tau_\ell(u_i,u_i)}{2} + \grad\bdot\bJ^{\mbox{\tiny{small}}}_\ell
= \Pi_\ell + \Lambda_\ell + \OL\tau_\ell(P,\grad\bdot\bu) 
- D^{\mbox{\tiny{small}}}_\ell + \epsilon^{\mbox{\tiny{small}}}_\ell,
\lb{smallKE}\ee
The SGS kinetic energy flux terms, $\Pi + \Lambda$
that appear as a sink in the large-scale budget (\ref{largeKE}) now appear as a source, representing the
energy gained by the small-scales from scales larger than $\ell$. 
$\bJ_\ell^{\mbox{\tiny{small}}}$ is spatial transport of $\wt{k}$, $\OL\tau_\ell(P,\grad\bdot\bu)$ is 
pressure dilatation taking place at scales $<\ell$,  $D_\ell^{\mbox{\tiny{small}}}$ is small-scale viscous dissipation,
and $\epsilon_\ell^{\mbox{\tiny{small}}}$ is energy input by external stirring that will vanish on average at scales $\ell \ll K_0^{-1}$ for a large-scale ${\bf F}^{<K_0}$ as we proved in Proposition \ref{Prop5}.  These terms are defined as
\begin{eqnarray}
&D^{\mbox{\tiny{small}}}_\ell(\bx)&  = \OL{\partial_j u_i \,\sigma_{ij}} -  \partial_j \wt{u}_i \,\OL{\sigma}_{ij} \\[0.3cm]
&J^{\mbox{\tiny{small}}}_j(\bx)&  =  \frac{\OL\rho\,\wt\tau(u_i,u_i)}{2}\wt{u}_j + \frac{1}{2}\OL\rho\,\wt\tau(u_i,u_i,u_j)
+\OL\tau(P,u_j) - (\OL{u_i\sigma_{ij}} - \wt{u}_i\OL{\sigma}_{ij}) \hspace{2cm}
\lb{smalltransport}   \\[0.4cm]
&\epsilon^{\mbox{\tiny{small}}}_\ell(\bx)& =  \OL\rho\,\wt\tau(u_i,F_i),\lb{smallinjectiondef}\\
\nonumber \end{eqnarray}
where $\wt\tau(f,g,h) = \wt{fgh} - \wt{f} \,\,\wt\tau(g,h) - \wt{g} \,\wt\tau(f,h) - \wt{h} \,\wt\tau(f,g) - \wt{f} \,\wt{g}\,\wt{h}$ in expression (\ref{smalltransport}) is a Favre-filter analogue of the ``generalized central moment'' introduced by Germano \cite{Germano92}.

\end{document}